\documentclass[11pt,a4 paper]{article}
\usepackage{}
\usepackage{cite}
\usepackage{amssymb}
\usepackage{amsmath}
\usepackage{epstopdf}
\usepackage[dvips]{graphicx}
\usepackage{caption}
\usepackage{subcaption}
\begin{document}
\thispagestyle{empty}

\noindent {\textbf{\Large  Evolution and thermodynamics of new holographic dark energy with bulk viscosity in modified $f(R, T)$ gravity }}\\

\vspace{0.2cm}

\noindent {\textbf{\small{C. P. Singh\footnote{Corresponding author}} and  {Milan Srivastava$^2$}}}\\

\vspace{0.2cm}

\noindent{ $^{1,2}$ Department of Applied Mathematics,\\
 Delhi Technological University,\\
 Bawana Road, Delhi-110 042, India.}\\
 \texttt{ $^1$cpsphd@rediffmail.com  \\
$^2$milandtu@gmail.com}\

\vspace{2cm}

\noindent \textbf{Abstract.} In this paper, bulk viscosity is introduced in new holographic dark energy (HDE) to describe the effects of cosmic non-perfect fluid on the evolution of the universe in modified $f(R,T)$ gravity. Assuming $f(R,T)=R+\lambda T$, where $R$ is the Ricci scalar, $T$, the trace of energy-momentum tensor and $\lambda$ is a constant, we derive a general function of Hubble parameter with bulk viscous form to provide a procedure for the viscous new HDE model building. Especially, we assume the total bulk viscosity coefficient proportional to the velocity of the expansion of the universe in such a way that $\zeta=\zeta_0+\zeta_1 H$. We obtain the solutions of the scale factor and deceleration parameter and classify all the possible scenarios (deceleration, acceleration and their transition) with different parameter regions chosen properly for positive and negative ranges of $\lambda$,  and $\zeta_0$ and $\zeta_1$ to analyze the evolution of the universe. It is observed that there is a transition from decelerated phase to accelerated phase at early or late time depending on the values of viscous terms. For large values of viscous terms it always accelerates through out the evolution. Furthermore, we also investigate the statefinder pair $\{r, s\}$ and $\textit{Om}$ diagnostics for the viscous new HDE model to discriminate with other existing DE models. The model evolution behaviors are shown in the planes of $r-s$, $r-q$ and $\textit{Om}-z$. It is found that behavior of trajectories in different planes depend on the bulk viscous coefficient. A small combination of $\zeta_0$ and $\zeta_1$ gives the quintessence like behavior whereas large combination of these two terms give Chaplygin gas like model. However, both the model approaches to $\Lambda CDM$ model in late time of evolution of the universe. Our viscous model is also different from other existing dark energy models. The evolution of effective equation of state parameter is -0.9745 which is very close to the observation. The entropy and generalized second law of thermodynamics are valid for this model under some constraints of bulk viscous coefficient. The analysis shows that the dark energy phenomena may be explained as the effect of bulk viscosity in the cosmic medium.\\

\noindent \textbf{Keywords:} Cosmology; Bulk viscosity; Dark energy; Modified $f(R,T)$ gravity.\\ \\ \\
\pagebreak

\tableofcontents

\section{Introduction}
\label{intro}
The observational studies including type Ia supernovae (SNeIa) \cite{riess,perlmutter}, cosmic microwave background radiation
(CMB)\cite{spe} and large scale structure (LSS) \cite{teg}, among others have confirmed that the present observable universe is in the phase of accelerated expansion. All these observations indicate the existence of a "dark energy" (DE) having negative pressure. The most important theoretical candidate for dark energy is the cosmological constant $\Lambda$ \cite{pad,sahni,peebles}, which fits the observations well, but it suffers from severe theoretical difficulties like fine -tuning and cosmic coincidence problems. Thus, dynamical dark energy models become popular, because they may alleviate the theoretical challenges faced by the $\Lambda CDM$ model. In cosmology, there are many phenomenological models of DE, {\it e.g.}, quintessence \cite{capo,gui,chiba,cal,feng,guo}, $k$-essence \cite{sch}, phantom energy \cite{calk,chi,vik}, quintom \cite{wei,zhao}, exotic Chaplygin gas \cite{set}, modified gravity theories \cite{noj,sot}, holographic dark energy \cite{li}, New agegraphic dark energy models \cite{cai}. Thus, dark energy has become one of the most important research areas in cosmology. However, our understanding about the nature of this energy is very modest, despite substantial progress both in theoretical and observational fields.\\
\indent In recent times, the holographic dark energy (HDE) is considered as a dynamics vacuum energy. It is stimulated by the holographic principle \cite{hooft,bou}, which says that the number of degrees of freedom of a physical system should scale with its bounding area rather than with its volume. It was suggested by Cohen {\it et al.} \cite{cohen} that in quantum field theory a short distance cutoff is related to a long distance cutoff due to the limit set by the formation of a black hole. They also suggested that the total energy in a region of size $L$ should not exceed the mass of a black hole of the same size. Under this assumption, Li \cite{li} proposed the HDE density as $\rho_{\Lambda}=3c^2M^{2}_{p}L^{-2}$, where $c^2$ is a dimensionless constant, $M^{-2}_{p}=8\pi G$ is the reduced Planck mass.\\
\indent Now, the problem is how to choose an appropriate infrared cutoff for the theory. Li \cite{li} has assumed Hubble length as the infrared cutoff ($\rho_d\propto H^2$) which resolves the fine-tuning problem but yields a wrong equation of state of DE,  $\omega_d=0$ \cite{hsu,xu}, which can not drive the cosmic acceleration. Later on, particle horizon as a different infrared cutoff was proposed which also gives $\omega_d>-1/3$. A suitable choice of infrared cutoff as a future event horizon was suggested by Li {\it et al.} \cite{lietal}, which explains the accelerated expansion of the universe. Gao {\it et al.} \cite{gao} has proposed infrared cutoff by replacing the future event horizon area with the inverse of the Ricci scalar curvature. They call this model the Ricci dark energy (RDE) model. Granda and Oliveros \cite{go} have proposed a new cutoff based on purely dimensional grounds, by adding a term involving the first derivative of the Hubble parameter. According to this new infrared cutoff, the energy density of HDE is given by
\begin{equation}
\rho_d=3M^{2}_{p}(\alpha H^2 +\beta \dot{H})
\end{equation}
\noindent where $H$ is the Hubble parameter and $\alpha$ and $\beta$ are constants which must satisfy the restrictions imposed by the current observational data. This is known as the new holographic dark energy (HDE). There are some works on the
new version of the holographic dark energy model \cite{m,o,k,j}. Therefore, it will be interesting to work with this new HDE model.\\
\indent It is observed that early inflation and recent expansion of the universe can be investigated through dissipative fluid. The role of the dissipative processes has been conscientiously studied \cite{barrow,brevik,odi,liu1} in the evolution of the universe. Viscosity is basically a measurement of a fluid's resistance to flow and is classified into two types, namely, bulk and shear viscosity. The general theory of dissipation in relativistic imperfect fluid was put on a firm foundation by Eckart \cite{eck}, and, in a somewhat different formulation by Landau and Lifshitz \cite{lan}. This is only the first order deviation from equilibrium and may has a causality problem, the full causal theory was developed by Israel and Stewart \cite{isr}, and has also been studied in the evolution of the early universe. However, because of the simple form of Eckart theory, it has been widely used by several authors to characterize the bulk viscous fluid.  The Eckart approach has been used in models explaining the recent acceleration of the universe with bulk viscous fluid. The idea of viscous DE models has been presented in different ways to understand the evolution of the universe. Many authors \cite{xu1,fengl,no,cap,cruz,meng,ren,hum,pan,singh1,singh2,singh3} have studied the DE phenomenon as an effect of the bulk viscosity in the cosmic medium. All these cited works are pioneer papers on bulk viscosity which show that for an appropriate viscosity coefficient, an accelerating cosmology can be achieved without the need of a cosmological constant \cite{gag}. At the late times, since we do not know the nature of the universe content very clearly, concern about the bulk viscosity is reasonable and practical. To our knowledge, such a possibility has been investigated only in the context of the primordial universe, concerning also the search of non-singular models. But many investigations show that the viscous pressure can play the role of an agent that drives the present acceleration of the universe. The motivation of the present work is to drive the present acceleration using the bulk viscous pressure in new HDE model.\\
\indent The modification in the geometrical part of Einstein-Hilbert action is very attractive way to resolve many problems in cosmology. The most famous modification of general relativity is the $f(R)$ gravity in which the Ricci scalar $R$ is replaced by a general function $f(R)$. This theory is consistent with the observations \cite{nod,ba}. Harko {\it et al.} \cite{har} proposed a new modified theory known as $f(R,T)$ gravity theory, where $R$ is the Ricci Scalar and $T$ stands for the trace of energy-momentum tensor. This modified theory presents a maximal coupling between geometry and matter. Many authors \cite{sha,cha,hark,singh4,baf} have studied modified $f(R,T)$ theory in different context to explain early and late time evolution of the universe. The new HDE model has not been yet discussed in detail in framework of $f(R,T)$ theory. Therefore, our aim is to study new HDE model with bulk viscosity in $f(R,T)$ gravity theory to explain the accelerated expansion of the universe.\\
\indent HDE is a great arena for modified gravity for a few reasons. The DE models (like $\Lambda CDM$) are alone able to explain the acceleration of the universe in the framework of the general relativity but it suffers some problems, like cosmological constant (the fine-tunning problem) and cosmic coincidence problems. To resolve these problems, either the left hand side of Einstein field equations, {\it i.e.}, geometrical part or right hand side, energy-momentum tensor has been modified.\\
\indent As a non-renormalizable theory, and currently the only known non-renormalizable theory, gravity is ultraviolet sensitive. In HDE the ultraviolet cutoff of the theory depends on the infrared cutoff. The infrared cutoff is set at the cosmological scale for the concern of dark energy. As a result, the ultraviolet cutoff is much affected. As we do not have a firm general relativity equation for quantum gravity, the best we can do is to take a modified gravity theory with the presence of HDE as a candidate of the infrared gravity theory. For this reason, actually, even if a modified theory is ruled out on earth, solar system or galactic scale experiments, the modified theory may still be considered together with HDE because cosmological scale is a completely different scale.\\
 \indent For many modified models to work as dark energy, they still have to solve the old cosmological constant problem and coincidence problem. HDE solves the problem for this modified theories. Thus, the modified theory in HDE can focus on the naturalness from first principle, dynamics of dark energy, agreement with observations, {\it etc}. Observationally, HDE has two parameters, which is relatively few compared with most dark energy models (although the cosmological constant has one parameter only). Thus, modified on top of HDE has stronger predictability compared to those on top of dark energy scenarios with more parameters (or even free functions). Modified gravity  theories have rich dynamics. It is thus interesting to study reconstructing those modified gravity theories using the dynamics of HDE. Different modified theories has different degree of freedoms and dynamics.\\
\indent The present paper is devoted to explore the effect of bulk viscosity in new HDE model within the context of modified $f(R,T)$ gravity. We consider the viscous new HDE model with pressureless matter. The time-dependent bulk viscous coefficient $\zeta=\zeta_0+\zeta_1 H$, where $\zeta_0$ and $\zeta_1$ are constants, has been assumed to discuss the evolution. We briefly discuss how the presence of viscous fluid could produce the late time acceleration. The viscous new HDE model gives time-dependent deceleration parameter which shows phase transition. We discuss the behavior of deceleration parameter by constraining on $\zeta_0$ and $\zeta_1$ and gravity parameter $\lambda$ which are summarize in tables. We also find the two independent geometrical diagnostics, namely statefinder pair and $\textit{Om}$ to discriminate the  new HDE model with other existing DE models. The effective equation of state parameter, entropy and generalized second law of thermodynamics are also discussed.\\
\indent This paper is organized in the following manner. In Section 2 the background and gravitational field equations of modified $f(R,T)$ gravity are written down. Section 3 deals with field equations for the viscous new HDE model in $f(R,T)$ gravity theory. Further, Section 3 is divided into subsections 3.1 -- 3.6 in which we find the solutions of the scale factor and the relevant physical and geometrical quantities like deceleration parameter, statefinder parameters, $\textit{Om}$ diagnostic, effective equation of state parameter, and entropy and generalized second law of thermodynamics, respectively, and discuss their evolutions in detail. We discuss and summarize our results in Section 4.\\
\indent In the present paper, we assume $8 \pi G = 1$ and $c = 1$.
\section{Review of Modified $f(R, T)$ gravity}
\label{sec:1}

The action of modified $f(R,T)$ gravity \cite{har} is defined as
\begin{equation}
S = \frac{1}{2} \int d^4 x \sqrt{-g} [f(R,T)+2 \mathcal{L}_{m}],
\end{equation}
where $f(R,T)$ is an arbitrary function of the Ricci scalar $R$ and trace $T$ of the stress energy tensor $T_{\mu\nu}$ of the matter, $g$ is the determinant of the metric tensor $g_{\mu\nu}$ and $\mathcal{L}_{m}$ represents the matter Lagrangian density. The stress energy tensor is defined as
\begin{equation}
T_{\mu\nu} = -\frac{2}{\sqrt{-g}}\frac{\delta (\sqrt{-g} \mathcal{L}_{m})}{\delta g^{\mu\nu}}.
\end{equation}
Assuming that the Lagrangian density $\mathcal{L}_{m}$ of matter depends only on the metric tensor $g_{\mu\nu}$ which leads to
\begin{equation}
T_{\mu\nu}=g_{\mu\nu}\mathcal{L}_{m}-2\frac{\delta \mathcal{L}_{m}}{\delta g_{\mu\nu}}.
\end{equation}
\noindent Variation of the action (2) with respect to the metric tensor $g_{\mu\nu}$ gives the field equations
\begin{equation}
f_R(R,T) R_{\mu\nu}-\frac{1}{2} f(R,T)g_{\mu\nu}+(g_{\mu\nu}\square-\triangledown_{\mu}\triangledown_{\nu})f_R(R,T)=T_{\mu\nu}-f_T(R,T)(T_{\mu\nu}+\circleddash_{\mu\nu}),
\end{equation}
\noindent where the tensor $\circleddash_{\mu\nu}$ is given by
\begin{equation}
\circleddash_{\mu\nu}=-2T_{\mu\nu}+g_{\mu\nu}\mathcal{L}_m-2g^{\alpha\beta}\frac{\partial^2\mathcal{L}_m}{\partial g^{\mu\nu}\partial g^{\alpha\beta}}.
\end{equation}
\noindent Here, $f_R=\partial f/\partial R$, $f_T=\partial f /\partial T$, $\square\equiv \triangledown^{\mu}\triangledown_{\mu}$, $\triangledown_{\mu}$ is the covariant derivative. \\
\indent Harko {\it et al.} \cite{har} introduced the form of $f(R,T)=R+f(T)$, where $f(T)=\lambda T$, $\lambda$ is an arbitrary constant. Using this form of $f(R,T)$, the field equations (5) are
\begin{equation}
R_{\mu\nu}-\frac{1}{2}R \;g_{\mu\nu}=T_{\mu\nu}- \lambda\;(T_{\mu\nu}+\circleddash_{\mu\nu})+ \frac{1}{2}\lambda\; g_{\mu\nu}\;T,
\end{equation}
\indent According to Harko {\it et al.} \cite{har} the matter Lagrangian $\mathcal{L}_{m}$ may be chosen as  $\mathcal{L}_{m} = -p$, where $p$ is the thermodynamical pressure of matter content of the universe. Then, Eq.(6) becomes as $\circleddash_{\mu\nu} = -2 T_{\mu\nu} - p\; g_{\mu\nu}$. Substituting this value into Eq.(7), we get
\begin{equation}
R_{\mu\nu} - \frac{1}{2} R\; g_{\mu\nu} = T_{\mu\nu} + \lambda\;(T_{\mu\nu} + p\; g_{\mu\nu}) + \frac{1}{2} \lambda T \;g_{\mu\nu}.
\end{equation}
\section{Field equations of bulk viscous new HDE model}
\label{sec:2}
We assume that the universe is filled with bulk viscous HDE fluid and dust dark matter(excluding the baryonic matter), and its geometry is given by a spatially flat Friedmann-Robertson-Walker (FRW) metric
\begin{equation}
ds^2 = dt^2 - a^2(t) \left[dr^2+r^2(d\theta^2+sin^2\theta d\phi^2)\right],
\end{equation}
where $a(t)$ is the scale factor, $t$ is the cosmic time and $(r, \theta, \phi)$ are the comoving coordinates.\\
\indent The stress-energy-momentum tensor in the presence of bulk viscous term is given by
\begin{equation}
T_{\mu\nu}=(\rho_m+\rho_d)u_{\mu}u_{\nu}+(g_{\mu\nu}+u_{\mu}u_{\nu}){\tilde{p_d}},
\end{equation}
\noindent where $\rho_m$ and $\rho_d$ are respectively the energy density of DM and new HDE, and ${\tilde{p_d}}$ is an effective pressure composed by the pressure $p_d$ of new HDE plus the bulk viscous pressure.\\
\indent Because of the assumed isotropic and homogeneity of the model, the shear viscosity plays no role, and only the bulk viscosity $\zeta$ has to be considered. In general, the presence of $\zeta$ does not have any influence upon the (00)-component of the equations of motion. The only change in the formalism because of viscosity is that the thermodynamical pressure  becomes replaced with the effective pressure ${\tilde{p_d}}$ defined as
\begin{equation}
{\tilde{p_d}}  = p_d - \zeta \triangledown_{\nu}u^{\nu},
\end{equation}
where $\zeta$ is a coefficient of bulk viscosity that arises in a fluid when it it out of local thermodynamic equilibrium and that induces a viscous pressure equals to $-\zeta \triangledown_{\nu}u^{\nu}$. The form of the above equation was originally proposed by Eckart \cite{eck} in the context of relativistic dissipative process occurring in thermodynamic systems went out of local thermal equilibrium. Many authors \cite{kre,cata,fab,hug} have used Eckart approach to explain the recent acceleration of the universe with bulk viscous fluid. In particular, it has been used to model bulk viscous dark fluids as responsible of the observed acceleration of the universe. It is assumed that the approximation of vanishing relaxation time is valid for this purpose for instance \cite{ave1,ave2,ave3}. In a paper \cite{ave1}, it is mentioned that Hiscock and Salmonson \cite{his} showed that a flat FRW model with a bulk viscous Boltzmann gas expands faster when the Eckart framework is used. These motivate us to use Eckart formalism, especially when one tries to look at the phenomenon of recent acceleration of the universe.\\
\indent In $f(R,T)$ theory when it is considered the effective pressure, the matter Lagrangian is $\mathcal{L}_{m} = - {\tilde{p_d}}$. Therefore,  Eq.(6) gives $\circleddash_{\mu\nu} = -2 T_{\mu\nu}- {\tilde{p_d}} \;g_{\mu\nu}$, where the trace, $T=g^{\mu\nu} T_{\mu\nu}$, has now the form of $T=\rho_m+\rho_d-3(p_d-\zeta \triangledown_{\nu}u^{\nu})$. \\
\indent Using above values of ${\tilde{p_d}}$, $\circleddash_{\mu\nu}$ and $T$, the field equations (8) for the line element (9) and energy stress tensor (10) yield
\begin{equation}
3 H^2 = \rho_m + \rho_d + \lambda\; (\rho_m +\rho_d + p_d -3 \zeta H) + \frac{1}{2} \lambda \;T,
\end{equation}
\begin{equation}
2 \dot H + 3 H^2 = -p_d + 3 \zeta H + \frac{1}{2} \lambda \;T.
\end{equation}
\noindent where $H=\dot{a}/{a}$ is the Hubble parameter and the bulk viscous pressure $-\zeta \triangledown_{\nu}u^{\nu}$ can be written as $-3\zeta H$. An over dot denotes time derivative.  We assume that a relation between $p_d$ and $\rho_d$ is connected by an equation of state (EoS), $p_d=\omega_d\; \rho_d$, where $\omega_d$ is the EoS parameter of new HDE. Therefore, from dynamical equations (12) and (13), a single evolution equation for $H$ can be obtained as
\begin{equation}
2 \dot H + (1+\lambda) [\rho_m + (1+\omega_d) \rho_d ] - 3(1+\lambda) \zeta H = 0.
\end{equation}
Following the Granda and Oliveros \cite{go}, the energy density of new HDE can be written as
\begin{equation}
\rho_d = 3 (\alpha H^2 + \beta \dot H),
\end{equation}
where $\alpha$ and $\beta$ are the dimensionless parameters, which must satisfy the restrictions imposed by the current observational data. \\
\indent Using (15) into (12), the energy density $\rho_m$ of DM can be obtained as
\begin{equation}
\rho_m = \frac{3}{(2+3\lambda)}\left[ (2-2\alpha - 3\lambda \alpha + \lambda \alpha \omega_d) H^2 -\lambda \zeta H -\beta (2 + 3\lambda -\lambda \omega_d) \dot H \right].
\end{equation}
Substituting (15) and (16) into (14), we get finally the first order differential equation for the Hubble parameter as
\begin{equation}
\dot H + (1+2\lambda\alpha\omega_d+\alpha\omega_d) A H^2 -(1+2\lambda)\zeta A H=0.
\end{equation}
where $A=\frac{3(1+\lambda)}{2+3\lambda+3\beta(1+\lambda)(1+2\lambda)\omega_d}$.\\

\indent On the thermodynamical grounds, $\zeta$ is conventionally chosen to be a positive quantity and generically depends on the cosmic time $t$, or redshift $z$, or the scale factor $a$, or the energy density $\rho_d$, or a more complicated combination form. Maartens \cite{maartens} assumed the bulk viscous coefficient as $\zeta \propto \rho^{n}$, where $n$ is a constant. Li and Barrow \cite{bli} have studied a unified model for the dark sectors with a single component universe consisting of bulk viscous dark matter. They have assumed the bulk viscous coefficient as a function of energy density alone. They have observed that the model shows an early deceleration and late time acceleration. In the Refs. \cite{meng,meng1,ren}, the most general form of bulk viscosity has been considered with generalized equation of state. Many authors have considered various forms of bulk viscous coefficient to discuss the cosmological models. Here, we analyze the properties of viscous new HDE by considering bulk viscous coefficient as \cite{meng,meng1,ren,ave1}
\begin{equation}
\zeta = \zeta_0 + \zeta_1 H,
\end{equation}
where $\zeta_0$ and $\zeta_1$ are positive constants. The motivation for choosing this bulk viscosity is that the viscosity phenomenon is associated with velocity and acceleration. We discuss the cosmological issues through a parameterized bulk viscosity that is a linear combination of two terms. The second term $\zeta_1 H$ describes bulk viscosity proportional to the Hubble parameter $H$, i.e., expansion ratio of the universe. The model with first term is discussed by many authors. Thus, a linear combination of these two terms may provide better results.\\
Using the ansatz (18) into (17), we get the evolution equation as
\begin{equation}
\dot H + \{1+(1+2\lambda)( \alpha \omega_d - \zeta_1)\} A H^2 -(1+2\lambda)\zeta_0 A H=0.
\end{equation}
In what follows we solve (19) to find the cosmological parameters, Hubble parameter, scale factor, deceleration parameter in terms of cosmic time $t$ and $a$ and discuss the behavior in detail.
\subsection{Evolution of the scale factor}
\label{sec:3}
On solving (19), we get
\begin{equation}
H = \frac{e^{(1+2\lambda)A\zeta_0 t}}{c_1 + \frac{\{1+(1+2\lambda)( \alpha \omega_d - \zeta_1)\}}{(1+2\lambda)\zeta_0}\;e^{(1+2\lambda)A\zeta_0 t}},
\end{equation}
where $c_1$ is the constant of integration. Now, using $H = {\dot a}/{a}$, we can obtained the scale factor as
\begin{equation}
a = c_2\left[ c_1 + \frac{\{1+(1+2\lambda)( \alpha \omega_d - \zeta_1)\}}{(1+2\lambda)\zeta_0} e^{(1+2\lambda)A\zeta_0 t}\right]^{\frac{1}{A\{1+(1+2\lambda)( \alpha \omega_d - \zeta_1)\}}},
\end{equation}
where $c_2$ is another integration constant.
Assuming $H=H_0$ at $t=t_0$, where $t_0$ is the present time when viscous new HDE starts to dominate, Eq.(20) can be written as
\begin{equation}
H(t) = H_0 e^{(1+2\lambda)A\zeta_0 (t-t_0)} \left[ 1+  \frac{\{1+(1+2\lambda)( \alpha \omega_d - \zeta_1)\} H_0}{(1+2\lambda)\zeta_0} \left(e^{(1+2\lambda)A\zeta_0 (t-t_0)}-1\right)\right]^{-1}.
\end{equation}
If we consider $a=a_0=1$ at $t=t_0$, then the scale factor (21) can be written as
\begin{equation}
a(t) = \left[ 1+  \frac{\{1+(1+2\lambda)( \alpha \omega_d -\zeta_1)\} H_0}{(1+2\lambda)\zeta_0} \left(e^{ (1+2\lambda)A\zeta_0 (t-t_0)}-1\right)\right]^{\frac{1}{A\{1+(1+2\lambda)( \alpha \omega_d - \zeta_1)\}}},
\end{equation}
where $\lambda\neq -1/2$ and $\zeta_0\neq 0$. It is observed that the scale factor is of  exponential form which can explain the phase transition. We can analyzed the behavior of the scale factor for all possible combinations of $(\zeta_0, \zeta_1)$ and model parameter $\lambda$. Taking $a(t)=0$, we obtain the cosmic time when the Big-Bang happens
\begin{equation}
t(\text{at Big-Bang})=t_0+\frac{1}{(1+2\lambda)A\zeta_0}\;ln \left[1-\frac{(1+2\lambda)\zeta_0}{\{1+(1+2\lambda)(\alpha \omega_d-\zeta_1)\}H_0}\right].
\end{equation}
In early time of evolution, the scale factor (23) can be approximated by
\begin{equation}
a\sim \left[  1+\frac{3H_0(1+\lambda)\{1+(1+2\lambda)(\alpha\omega_d-\zeta_1)\}}
{2+3\lambda+3\beta(1+\lambda)(1+2\lambda)}(t-t_0) \right]^{\frac{1}{A\{1+(1+2\lambda)( \alpha \omega_d - \zeta_1)\}}},
\end{equation}
which shows the decelerated expansion in early time. In late time of evolution, the scale factor behaves as
\begin{equation}
a(t)\sim exp\left[  (1+2\lambda)\zeta_0 A (t-t_0) \right],
\end{equation}
which shows the de Sitter universe, {\it i.e.}, the universe expands with accelerated rate in the late-time of the evolution. This shows that the scale factor at the respective limits has an earlier decelerated phase followed by an accelerated phase in the later stage of the evolution.\\
\indent Let us compute the second order derivatives of (23), which is given by
\begin{eqnarray}
\frac{d^2a}{dt^2}&=&H_0^2 \Big(e^{2(1+2\lambda)\zeta_0A(t-t_0)}\Big) \Bigg[ 1+\frac{[(1+2\lambda)\zeta_0-\{1+(1+2\lambda)(\alpha\omega_d-\zeta_1)\}H_0]A}{H_0}e^{-(1+2\lambda)\zeta_0A(t-t_0)} \Bigg]\nonumber\\&& \times\Bigg[ 1+\frac{\{1+(1+2\lambda)(\alpha\omega_d-\zeta_1)\}H_0}{(1+2\lambda)\zeta_0}\Big(e^{(1+2\lambda)\zeta_0A(t-t_0)}-1 \Big) \Bigg]^{\frac{1}{A\{1+(1+2\lambda)(\alpha\omega_d-\zeta_1)\}}-2}.
\end{eqnarray}
\noindent Equating (27) to zero to get the transition time, ${\it t_{\;trans}}$ between the decelerated to the accelerated expansion epochs, which is given by
\begin{equation}
t_{\;trans}=t_0+\frac{1}{A(1+2\lambda)\zeta_0}\;ln\Bigg\{\frac{A[\{1+(1+2\lambda)
(\alpha\omega_d-\zeta_1)\}H_0-(1+2\lambda)\zeta_0]}{H_0}\Bigg\}.
\end{equation}
\noindent Using (23) into (22), the Hubble parameter in terms of the scale factor can be written as
\begin{equation}
H(a)= \frac{(1+2\lambda)H_0}{[1+(1+2\lambda)(\alpha\omega_d-\zeta_1)]}\left[ \frac{\zeta_0}{H_0}+\left\{ \frac{[1+(1+2\lambda)(\alpha\omega_d-\zeta_1)]}{(1+2\lambda)}-\frac{\zeta_0}{H_0} \right\}a^{-A\{1+(1+2\lambda)(\alpha\omega_d-\zeta_1)\}}\right].
\end{equation}
Differentiating (29) with respect to $a$, we obtain
\begin{eqnarray}
\frac{d\dot a}{da}&=&  \frac{(1+2\lambda) H_0}{[1+(1+2\lambda)(\alpha\omega_d-\zeta_1)]} \Bigg[ \frac{\zeta_0}{H_0}-\left( \frac{[1+(1+2\lambda)(\alpha\omega_d-\zeta_1)]}{(1+2\lambda)}-\frac{\zeta_0}{H_0} \right)\nonumber\\&&\times \left( \frac{1+3(1+\lambda)(1+2\lambda)\{(\alpha-\beta)\omega_d-\zeta_1\}}
{2+3\lambda+3\beta(1+\lambda)(1+2\lambda)\omega_d} \right) a^{-A\{1+(1+2\lambda)( \alpha \omega_d - \zeta_1)\}}\Bigg].
\end{eqnarray}
Equating (30) to zero, the transition between the decelerated to the accelerated phase in terms of the scale factor can be written as
\begin{eqnarray}
a_T &=& \Bigg[ \Bigg\{ \frac{1+3(1+\lambda)(1+2\lambda)[(\alpha-\beta)\omega_d-\zeta_1]}{(1+2\lambda)
\{2+3\lambda+3\beta(1+\lambda)(1+2\lambda)\omega_d \}\zeta_0}\Bigg\}\nonumber\\&&\times\Big\{ [1+(1+2\lambda)(\alpha\omega_d-\zeta_1)]H_0-(1+2\lambda)\zeta_0 \Big\} \Bigg]^{\frac{1}{
A[1+(1+2\lambda)(\alpha\omega_d-\zeta_1)]}},
\end{eqnarray}
and the corresponding transition redshift $z=a^{-1}-1$ is
\begin{eqnarray}
z_T = &=& \Bigg[ \Bigg\{ \frac{1+3(1+\lambda)(1+2\lambda)[(\alpha-\beta)\omega_d-\zeta_1]}
{(1+2\lambda)\{2+3\lambda+3\beta(1+\lambda)(1+2\lambda)\omega_d \}\zeta_0}\Bigg\}\nonumber\\&&\times\Big\{ [1+(1+2\lambda)(\alpha\omega_d-\zeta_1)]H_0-(1+2\lambda)\zeta_0 \Big\} \Bigg]^{-\frac{1}{A[1+(1+2\lambda)(\alpha\omega_d-\zeta_1)]}}-1.
\end{eqnarray}
{\begin{center}
\begin{tabular}{c}
\begin{minipage}{220pt}
\includegraphics[width=220pt]{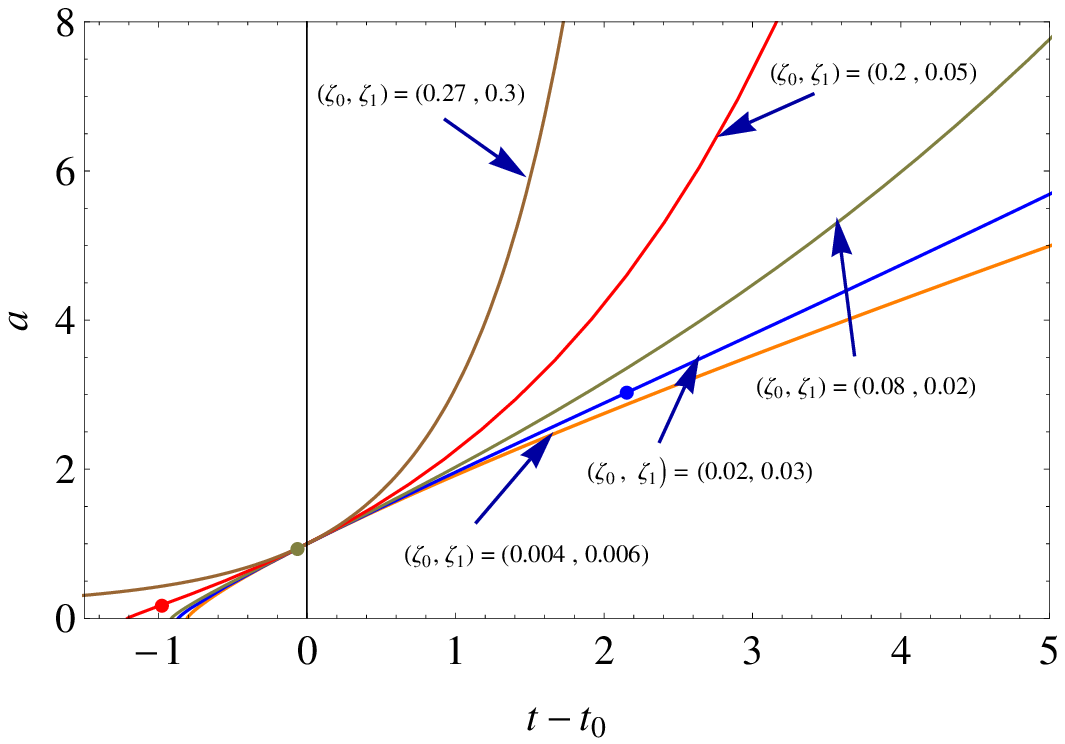}
\end{minipage}
\end{tabular}
\end{center}
{\footnotesize \textbf{Fig.1} The evolution of scale factor with respect to $(t-t_0)$ for different combination of $\zeta_0$ and $\zeta_1$ with $\omega_d=-0.5$, $\lambda=0.06$ $H_0=1$, $\alpha=0.8502$ and $\beta=0.4817$.} \\

Here, $T$ stands for transition. From (31) and (32), we observe that the transition from decelerated to accelerated epoch takes place at present time, {\it i.e.}, at $a_T=1$ or $z_T=0$ for $(\zeta_0+ H_0\zeta_1) = \frac{\{1+3(1+\lambda) (1+2\lambda)(\alpha-\beta) \omega_d\}H_0}{3(1+\lambda)(1+2\lambda)}$ depending on positive or negative value of $\lambda$. Especially, substituting the observational values of model parameters $\alpha=0.8502$ and $\beta=0.4817$ \cite{wangxu} along with $\omega_d=-0.5$, $H_0=1$, we get $\zeta_0+\zeta_1=0.096$ for positive value of $\lambda=0.06$ and $\zeta_0+\zeta_1=0.218$ for negative value of $\lambda=-0.06$, respectively. Thus, for the present time transition from deceleration to acceleration takes place at $\zeta_0+\zeta_1=0.096$ for positive values of $\lambda=0.06$ and for negative value of $\lambda=-0.06$, we get $\zeta_0+\zeta_1=0.218$. A plot of the evolution of the scale factor is given in Fig. 1. for different values of a combination of $\zeta_0$ and $\zeta_1$ and positive value of $\lambda$. For  $0<(\zeta_0+\zeta_1)< 0.096$, the scale factor has a deceleration phase followed by an accelerated phase in late time. For $(\zeta_0+\zeta_1)=0.096$, the transition takes place at present time and for $(\zeta_0+\zeta_1)>0.096$, the transition from decelerated phase to accelerated phase occurs at early time. For higher combination of $(\zeta_0,\zeta_1)$, acceleration takes place in infinite past. Similar behavior can be observed for the negative value of $\lambda$. Fig.1 plots the graph of the scale factor versus time for different combinations of $(\zeta_0,\zeta_1)$. A dot on each trajectory denotes the present time transition from decelerated phase to accelerated phase.
\subsection{Behavior of the deceleration parameter}
\label{sec:4}
\indent Now, the evolutionary behavior of the universe can also be discussed by the deceleration parameter, which is defined by $q = -\frac{a \ddot a}{\dot a^2}$. Using (23), the deceleration parameter can be obtained as
\begin{equation}
q(t) = \Big\{\frac{[\{1+(1+2\lambda) (\alpha \omega_d -\zeta_1)\} H_0 - (1+2\lambda)\zeta_0]A}{ H_0}\Big\}e^{-(1+2\lambda)\zeta_0 A(t-t_0)} - 1.
\end{equation}
\indent Equation (33) shows that the deceleration parameter is time-dependent which may describe the phase transition. It can be observed that deceleration parameter must change its sign at $t=t_0$. The sign of $q$ is positive for $t<t_0$ and it is negative for $t>t_0$. \\
\indent The deceleration parameter in terms of $a$ can be obtained as
\begin{align}
q(a)&=A[\{1+(1+2\lambda)(\alpha\omega_d-\zeta_1)\}H_0-(1+2\lambda)
\zeta_0]\notag\\ \quad{}{}& \times\left[ \frac{[1+(1+2\lambda)(\alpha\omega_d-\zeta_1)]}{(1+2\lambda)\zeta_0
\left\{a^{A[1+(1+2\lambda)(\alpha\omega_d-\zeta_1)]}-1\right\}
+[1+(1+2\lambda)(\alpha\omega_d-\zeta_1)]H_0} \right]-1.
\end{align}
In the terms of redshift, the deceleration parameter is given by
\begin{align}
q(z)&=A[\{1+(1+2\lambda)(\alpha\omega_d-\zeta_1)\}H_0-(1+2\lambda)
\zeta_0]\notag\\ \quad{}{}& \times\left[ \frac{[1+(1+2\lambda)(\alpha\omega_d-\zeta_1)]}{(1+2\lambda)\zeta_0\left\{ (1+z)^{-A[1+(1+2\lambda)(\alpha\omega_d-\zeta_1)]}-1 \right\}+[1+(1+2\lambda)(\alpha\omega_d-\zeta_1)]H_0} \right]-1.
\end{align}
The present value of $q$ corresponds to $z=0$ or $a=1$ is,
\begin{equation}
q_0=\frac{[\{1+(1+2\lambda)(\alpha\omega_d-\zeta_1)\}H_0-(1+2\lambda)
\zeta_0]A}{H_0}-1.
\end{equation}
This equation shows that if $(\zeta_0+ H_0\zeta_1) = \frac{\{1+3(1+\lambda) (1+2\lambda)(\alpha-\beta) \omega_d\}H_0}{3(1+\lambda)(1+2\lambda)}$, the value of $q_0=0$. Thus, the transition into accelerating phase would occur at present time for this combination of $(\zeta_0,\zeta_1)$. Especially, taking $\alpha=0.8502$, $\beta=0.4817$, $\omega_d=-0.5$, $H_0=1$ and $\lambda=0.06$ in above expression, we get $\zeta_0+\zeta_1=0.096$, which gives $q_0=0$.\\
\indent Tables 1-3 discuss the behavior of deceleration parameter and corresponding evolution for different ranges of $\lambda$ under constraints on $(\zeta_0+H_0\zeta_1)$. We consider three different values of $\omega_d$ of three different phases, {\it e.g.}, $\omega_d = -0.5, \;\;-1\;\; \text{and} \;\;-1.1414$, respectively. We assume the observational values of the model parameters are $\alpha=0.8502$ and $\beta=0.4817$ \cite{wangxu}.\\
\begin{center}
\begin{table*}
\small
\caption{Variation of $q$ for $\omega_d = -0.5$, $\alpha=0.8502$, $\beta=0.4817$}
\label{tab:1}       
\begin{tabular}{llll}
\hline\noalign{\smallskip}
$\lambda$ & Constraints on $\zeta_0$ and $\zeta_1$ & $q$ & Evolution of the Universe  \\
\noalign{\smallskip}\hline\noalign{\smallskip}
$\lambda\ge 1.272$ & For all $\zeta_0>0$ and $\zeta_1>0$ & Positive & Decelerated expansion\\
$0.24\le \lambda< 1.272$ & For all $\zeta_0>0$ and $\zeta_1>0$ & Negative & Accelerated expansion \\
$-0.5<\lambda<0.24$ & $0<(\zeta_0+H_0 \zeta_1) <\frac{\{1-0.55275(1+\lambda)(1+2\lambda)\}H_0}{3(1+\lambda)(1+2\lambda)}$ & $+$ve to $-$ve & Transition from dec. to acc.\\
  & $(\zeta_0+H_0 \zeta_1)\ge\frac{\{1-0.55275(1+\lambda)(1+2\lambda)\}H_0}{3(1+\lambda)(1+2\lambda)}$ & Negative & Accelerated expansion\\
$-0.69\le \lambda \le -0.5$ & For all $\zeta_0>0$ and $\zeta_1>0$ & Positive & Decelerated expansion\\
$-1\le \lambda < -0.69$ & For all $\zeta_0>0$ and $\zeta_1>0$ & Negative & Accelerated expansion\\
$-1.73\le \lambda <-1$ & $0<(\zeta_0+H_0 \zeta_1) < \frac{\{1-0.55275(1+\lambda)(1+2\lambda)\}H_0}{3(1+\lambda)(1+2\lambda)}$ & $-$ve to $+$ve & Transition from acc. to dec.\\
 & $(\zeta_0+H_0 \zeta_1) \ge \frac{\{1-0.55275(1+\lambda)(1+2\lambda)\}H_0}{3(1+\lambda)(1+2\lambda)}$ & Positive & Decelerated expansion\\
$\lambda<-1.73$ & For all $\zeta_0>0$ and $\zeta_1>0$ & Positive & Decelerated expansion\\
\noalign{\smallskip}\hline
\end{tabular}
\end{table*}
\end{center}
\begin{center}
\begin{table*}
\small
\caption{Variation of $q$ for $\omega_d = -1$, $\alpha=0.8502$, $\beta=0.4817$}
\label{tab:1}       
\begin{tabular}{llll}
\hline\noalign{\smallskip}
$\lambda$ & Constraints on $\zeta_0$ and $\zeta_1$ & $q$ & Evolution of the Universe  \\
\noalign{\smallskip}\hline\noalign{\smallskip}
$\lambda\ge 0.265$ & For all $\zeta_0>0$ and $\zeta_1>0$ & Positive & Decelerated expansion\\
$-0.032\le \lambda< 0.265$ & For all $\zeta_0>0$ and $\zeta_1>0$ & Negative & Accelerated expansion \\
$-0.5<\lambda<-0.032$ & $0<(\zeta_0+H_0 \zeta_1) <\frac{\{1-1.1055(1+\lambda)(1+2\lambda)\}H_0}{3(1+\lambda)(1+2\lambda)}$ & $+$ve to $-$ve & Transition from dec. to acc.\\
  & $(\zeta_0+H_0 \zeta_1) \ge\frac{\{1-1.1055(1+\lambda)(1+2\lambda)\}H_0}{3(1+\lambda)(1+2\lambda)}$ & Negative & Accelerated expansion\\
$-0.727< \lambda \le -0.5$ & For all $\zeta_0>0$ and $\zeta_1>0$ & Positive & Decelerated expansion\\
$-1\le \lambda \le -0.727$ & For all $\zeta_0>0$ and $\zeta_1>0$ & Negative & Accelerated expansion\\
$-1.46\le \lambda <-1$ & $0<(\zeta_0+H_0 \zeta_1) < \frac{\{1-1.1055(1+\lambda)(1+2\lambda)\}H_0}{3(1+\lambda)(1+2\lambda)}$ & $-$ve to $+$ve & Transition from acc. to dec.\\
 & $(\zeta_0 +H_0 \zeta_1) \ge \frac{\{1-1.1055(1+\lambda)(1+2\lambda)\}H_0}{3(1+\lambda)(1+2\lambda)}$ & Positive & Decelerated expansion\\
$\lambda<-1.46$ & For all $\zeta_0>0$ and $\zeta_1>0$ & Positive & Decelerated expansion\\
\noalign{\smallskip}\hline
\end{tabular}
\end{table*}
\end{center}
\begin{center}
\begin{table*}
\small
\caption{Variation of $q$ for $\omega_d = -1.1414$, $\alpha=0.8502$, $\beta=0.4817$}
\label{tab:1}       
\begin{tabular}{llll}
\hline\noalign{\smallskip}
$\lambda$ & Constraints on $\zeta_0$ and $\zeta_1$ & $q$ & Evolution of the Universe  \\
\noalign{\smallskip}\hline\noalign{\smallskip}
$\lambda\ge 0.145$ & For all $\zeta_0>0$ and $\zeta_1>0$ & Positive & Decelerated expansion\\
$-0.072\le \lambda< 0.145$ & For all $\zeta_0>0$ and $\zeta_1>0$ & Negative & Accelerated expansion \\
$-0.5<\lambda<-0.072$ & $0<(\zeta_0+H_0 \zeta_1) <\frac{\{1-1.26182(1+\lambda)(1+2\lambda)\}H_0}{3(1+\lambda)(1+2\lambda)}$ & $+$ve to $-$ve & Transition from dec. to acc.\\
  & $(\zeta_0+H_0 \zeta_1) \ge\frac{\{1-1.26182(1+\lambda)(1+2\lambda)\}H_0}{3(1+\lambda)(1+2\lambda)}$ & Negative & Accelerated expansion\\
$-0.735\le \lambda \le -0.5$ & For all $\zeta_0>0$  and $\zeta_1>0$ & Positive & Decelerated expansion\\
$-1\le \lambda < -0.735$ & For all $\zeta_0>0$ and $\zeta_1>0$ & Negative & Accelerated expansion\\
$-1.42\le \lambda <-1$ & $0<(\zeta_0+H_0 \zeta_1) < \frac{\{1-1.26182(1+\lambda)(1+2\lambda)\}H_0}{3(1+\lambda)(1+2\lambda)}$ & $-$ve to $+$ve & Transition from acc. to dec.\\
 & $(\zeta_0+H_0 \zeta_1) \ge \frac{\{1-1.26182(1+\lambda)(1+2\lambda)\}H_0}{3(1+\lambda)(1+2\lambda)}$ & Positive & Decelerated expansion\\
$\lambda<-1.42$ & For all $\zeta_0>0$ and $\zeta_1>0$ & Positive & Decelerated expansion\\
\noalign{\smallskip}\hline
\end{tabular}
\end{table*}
\end{center}
\indent In table 1, we observe that for any positive value of $\zeta_0$ and $\zeta_1$, the model corresponds to the decelerated expansion throughout the evolution for $\lambda\ge 1.272$,\; $-0.69\le\lambda\le -0.5$ and $\lambda<-1.73$, and the model corresponds to the accelerated expansion throughout the evolution for $0.24\le \lambda < 1.272$ and $-1\le \lambda <-0.69$. For smaller values of $\zeta_0$ and $\zeta_1$, {\it i.e.}, $0<(\zeta_0+H_0 \zeta_1)<\frac{\{1-0.55275(1+\lambda)(1+2\lambda)\}H_0}{3(1+\lambda)(1+2\lambda)}$, the universe shows the phase transition from positive to negative for $-0.5<\lambda<0.24$ and  shows the phase transition from negative to positive for $-1.73\le \lambda<-1$. The larger values of $\zeta_0$ and $\zeta_1$, {\it i.e.}, $(\zeta_0+H_0 \zeta_1)\ge\frac{\{1-0.55275(1+\lambda)(1+2\lambda)\}H_0}{3(1+\lambda)(1+2\lambda)}$, shows that the universe represents the accelerated expansion throughout the evolution for $-0.5<\lambda<0.24$ and represents the decelerated expansion throughout the evolution for $-1.73\le \lambda<-1$.\\
\indent Table 2 represents the analysis of the evolution of the universe for $\omega_d=-1$ along with the observational values of model parameters. We observe that for any positive value of $\zeta_0$ and $\zeta_1$, the model corresponds to the decelerated expansion throughout the evolution for $\lambda\ge 0.265$, $-0.727<\lambda\le -0.5$ and $\lambda<-1.46$ ranges of $\lambda$, and the model corresponds to the accelerated expansion throughout the evolution for $-0.032\le \lambda < 0.265$ and $-1\le \lambda \le -0.727$.  For smaller values of $\zeta_0$ and $\zeta_1$, {\it i.e.}, $0<(\zeta_0+H_0 \zeta_1)<\frac{\{1-1.1055(1+\lambda)(1+2\lambda)\}H_0}{3(1+\lambda)(1+2\lambda)}$, the universe shows the phase transition from deceleration to acceleration for $-0.5<\lambda<-0.032$ and shows the phase transition from acceleration to deceleration for $-1.46\le \lambda<-1$. The larger values of $\zeta_0$ and $\zeta_1$, {\it i.e.}, $(\zeta_0+H_0 \zeta_1)\ge\frac{\{1-1.1055(1+\lambda)(1+2\lambda)\}H_0}{3(1+\lambda)(1+2\lambda)}$, shows that the universe accelerates throughout the evolution for $-0.5<\lambda<-0.032$ and decelerates throughout the evolution for $-1.46\le \lambda<-1$.\\
\indent In Table 3, we consider the value of $\omega_d<-1$, {\it e.g.}, $\omega_d=-1.1414$ with the same observational values of model parameters. We observe that for any positive value of $\zeta_0$ and $\zeta_1$, the model shows the decelerated expansion throughout the evolution for $\lambda\ge 0.145$, $-0.735\le\lambda\le -0.5$ and $\lambda<-1.42$, and the model corresponds to the accelerated expansion throughout the evolution for $-0.072\le \lambda < 0.145$ and $-1\le \lambda <-0.735$. The smaller values of $\zeta_0$ and $\zeta_1$ , {\it i.e.}, $0<(\zeta_0 +H_0 \zeta_1)<\frac{\{1-1.26182(1+\lambda)(1+2\lambda)\}H_0}{3(1+\lambda)(1+2\lambda)}$, shows the phase transition from deceleration to acceleration for $-0.5<\lambda<-0.072$ and the phase transition from acceleration to deceleration occur for $-1.42\le \lambda<-1$. For the larger values of $\zeta_0$ and $\zeta_1$, {\it i.e.}, $(\zeta_0+H_0 \zeta_1)\ge\frac{\{1-1.26182(1+\lambda)(1+2\lambda)\}H_0}{3(1+\lambda)(1+2\lambda)}$, the universe represents the accelerated expansion throughout the evolution for $-0.5<\lambda<-0.072$ and shows the decelerated expansion throughout the evolution for $-1.42\le \lambda<-1$.

\subsection{Statefinder diagnostic}
\label{6}
Sahni {\it et al.} \cite{sahni1} and Alam {\it et al.} \cite{alam} proposed a new geometrical diagnostic pair $\{r,s\}$ called statefinder. The statefinder is a geometrical diagnostic in the sense that it is constructed from a space-time metric directly, and allows us to characterise the properties of DE in a model independent manner. It is dimensionless and is constructed from the scale factor $a(t)$ and its derivatives up to the third order. The statefinder pair is defined as
\begin{equation}
r=\frac{\dddot a}{a H^3}\;\;\; \text{and} \;\;\; s= \frac{r-1}{3(q-1/2)}.
\end{equation}
This pair has a fixed point values $\{r,s\}=\{1,0\}$ and $\{r,s\}=\{1,1\}$ for $\Lambda CDM$ and $SCDM$ models, respectively. The statefinder method has been extensively used in the literature to distinguish among various models of dark energy and modified theories of gravity. \\
\indent On substituting the required values in above, we get
\begin{eqnarray}
r&=&1+\left[\frac{3A\Big\{1-\frac{A[(1+(1+2\lambda)(\alpha\omega_d
-\zeta_1)]}{3}\Big\}\Big\{(1+2\lambda)\zeta_0-[(1+(1+2\lambda)(\alpha\omega_d
-\zeta_1)]H_0\Big\}}{H_0\;e^{(1+2\lambda)\zeta_0A(t-t_0)}}\right]
\nonumber\\&&+\;\;\left[ \frac{A^2\Big\{(1+2\lambda)\zeta_0-[(1+(1+2\lambda)(\alpha\omega_d
-\zeta_1)]H_0\Big\}^2}{H_0^2\;e^{2(1+2\lambda)\zeta_0A(t-t_0)}} \right],\nonumber\\&&
\end{eqnarray}
and
\begin{equation}
s=\frac{\frac{2A\Big\{1-\frac{A[(1+(1+2\lambda)(\alpha\omega_d
-\zeta_1)]}{3}\Big\}\Big\{(1+2\lambda)\zeta_0-[(1+(1+2\lambda)(\alpha\omega_d
-\zeta_1)]H_0\Big\}}{3H_0e^{(1+2\lambda)\zeta_0A(t-t_0)}} +\frac{2A^2\Big\{(1+2\lambda)\zeta_0-[(1+(1+2\lambda)(\alpha\omega_d
-\zeta_1)]H_0\Big\}^2}{9H_0^2e^{2(1+2\lambda)\zeta_0A(t-t_0)}}}
{\frac{2A\Big\{[(1+(1+2\lambda)(\alpha\omega_d-\zeta_1)]H_0-(1+2\lambda)\zeta_0\Big\}}
{3H_0e^{(1+2\lambda)\zeta_0A(t-t_0)}}-1},
\end{equation}
which are time dependent due to the presence of bulk viscosity coefficient. Equations (38) and (39) show that in the limit $(t-t_0)\rightarrow \infty $, the statefinder parameters $\{r,s\}\rightarrow \{1,0\}$, a value corresponding to the $\Lambda CDM$ model. Hence, the viscous new HDE model resembles the $\Lambda CDM$ model in future.  Now, we can plot the $r-s$ trajectory in $r-s$ plane and $r-q$ trajectory in $r-q$ plane to analyse viscous new HDE model in the framework of $f(R,T)$ theory. It can be observed that the values of $\{r,s\}$ depend on the choice of coupling parameter $\lambda$ and the viscosity coefficients ($\zeta_0, \zeta_1$). We consider the observational values of model parameters $\alpha=0.8502$, $\beta=0.4817$ and $H_0=1$, $t_0=1$, $\omega_d=-0.5$ to plot these trajectories for positive and negative value of $\lambda$ along with the different combinations of ($\zeta_0, \zeta_1$). The $r-s$ and $r-q$ trajectories for different combinations of ($\zeta_0, \zeta_1$) and positive value of $\lambda$ ({\it e.g.}, $\lambda=0.06$) are shown in Figs. 2a and 2b, respectively. Figs. 3a and 3b show the respective $r-s$ and $r-q$ trajectories for different combinations of ($\zeta_0, \zeta_1$) and negative value of $\lambda$ ({\it e.g.}, $\lambda=-0.06$). The present value of the statefinder pair are
\begin{eqnarray}
r_0&=&1+\left[\frac{3A\Big\{1-\frac{A[(1+(1+2\lambda)(\alpha\omega_d
-\zeta_1)]}{3}\Big\}\Big\{(1+2\lambda)\zeta_0-[(1+(1+2\lambda)(\alpha\omega_d
-\zeta_1)]H_0\Big\}}{H_0}\right]
\nonumber\\&&+\;\;\left[ \frac{A^2\Big\{(1+2\lambda)\zeta_0-[(1+(1+2\lambda)(\alpha\omega_d
-\zeta_1)]H_0\Big\}^2}{H_0^2} \right],\nonumber\\&&
\end{eqnarray}
and
\begin{equation}
s_0=\frac{\frac{2A\Big\{1-\frac{A[(1+(1+2\lambda)(\alpha\omega_d
-\zeta_1)]}{3}\Big\}\Big\{(1+2\lambda)\zeta_0-[(1+(1+2\lambda)(\alpha\omega_d
-\zeta_1)]H_0\Big\}}{3H_0} +\frac{2A^2\Big\{(1+2\lambda)\zeta_0-[(1+(1+2\lambda)(\alpha\omega_d
-\zeta_1)]H_0\Big\}^2}{9H_0^2}}
{\frac{2A\Big\{[(1+(1+2\lambda)(\alpha\omega_d-\zeta_1)]H_0-(1+2\lambda)\zeta_0\Big\}}
{3H_0}-1}.
\end{equation}
{\begin{center}
\begin{tabular}{cc}
\begin{minipage}{185pt}
\includegraphics[width=185pt]{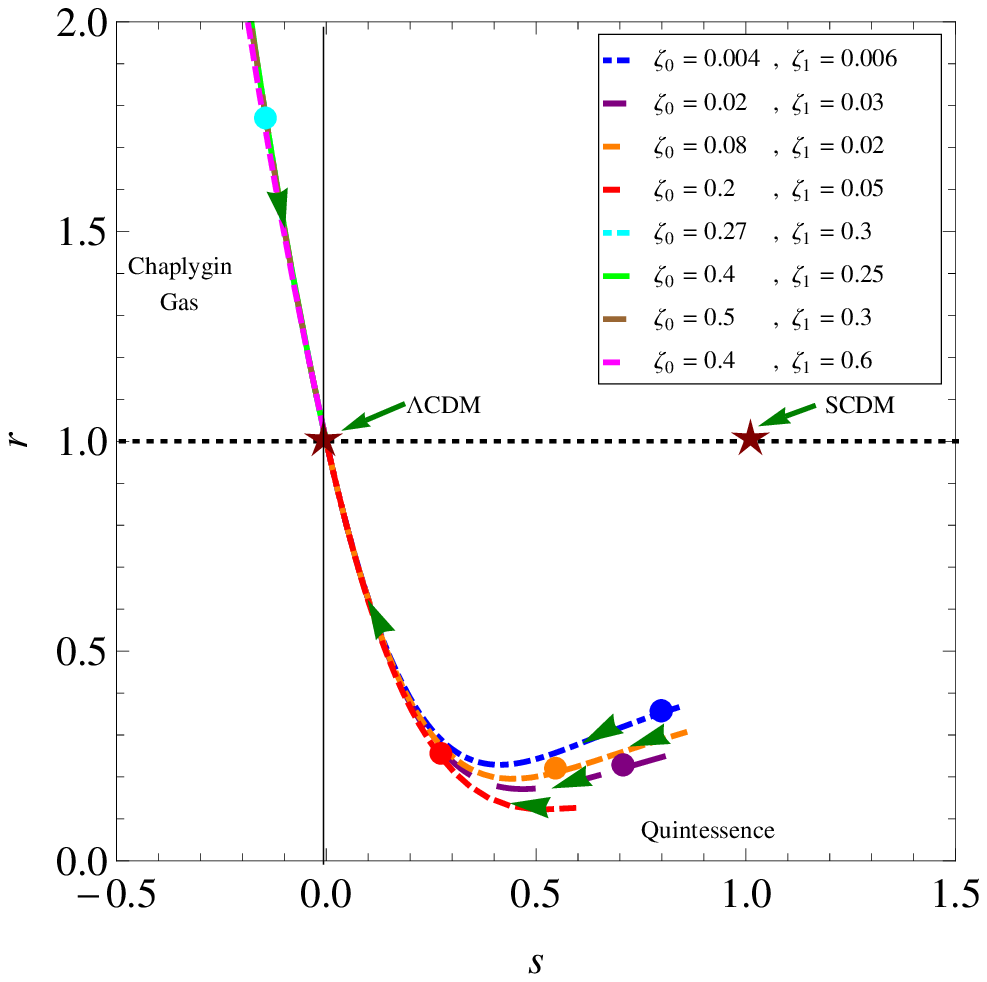}
{\footnotesize \textbf{Fig.2(a)} The $r-s$ trajectories are plotted in $r-s$ plane for different combinations of ($\zeta_0,\zeta_1$) with $\omega_d=-0.5$, $\alpha=0.8502$, $\beta=0.4817$ and $\lambda=0.06$. The arrows represent the directions of the evolutions of statefinder diagnostic pair with time.}
\end{minipage}&\begin{minipage}{185pt}
\includegraphics[width=210pt]{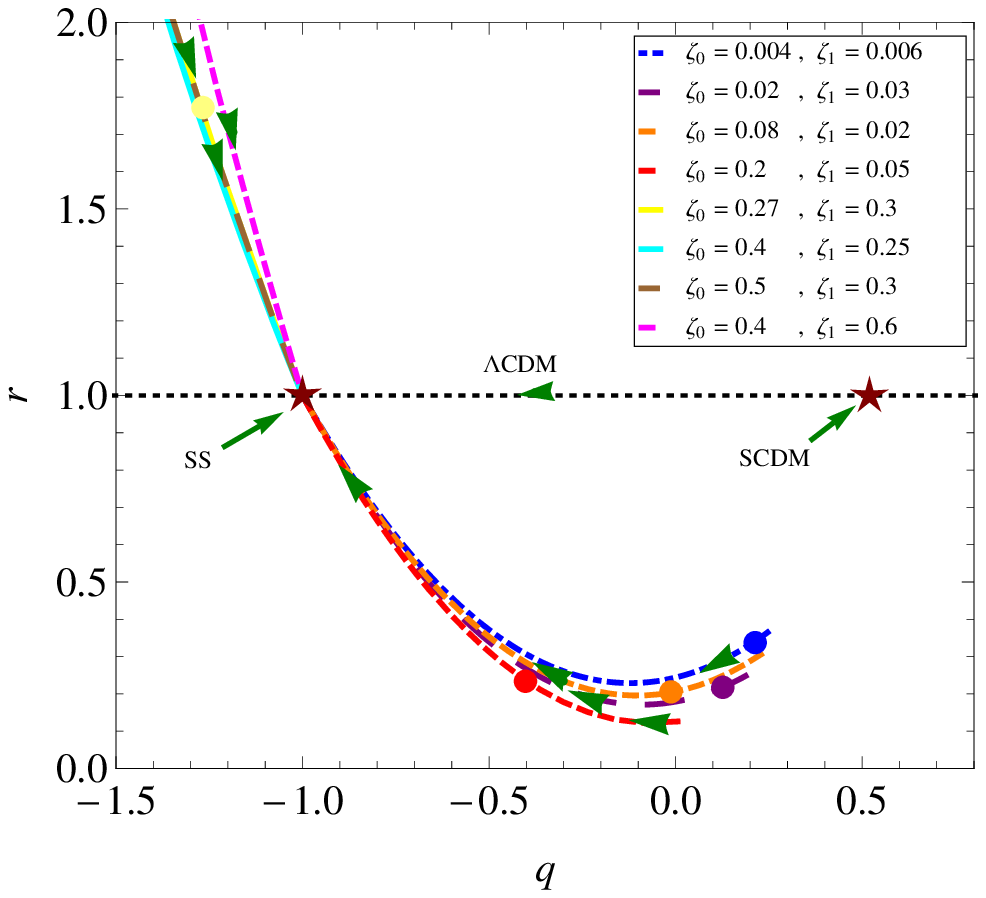}
{\footnotesize \textbf{Fig.2(b)} The $r-q$ trajectories are plotted in $r-q$ plane for different combinations of ($\zeta_0,\zeta_1$) with $\omega_d=-0.5$, $\alpha=0.8502$, $\beta=0.4817$ and $\lambda=0.06$. The arrows represent the directions of the evolutions of statefinder diagnostic pair with time.}
\end{minipage}
\end{tabular}
\end{center}
\indent In Figs. 2a and 3a, stars represent the fixed point values of $\Lambda CDM$ and $SCDM$ models, dots represent the present time values of $\{r,s\}=\{r_0,s_0\}$ and $\{r,q\}=\{r_0,q_0\}$, and the arrows represent the direction of the trajectories. The $r-s$ planes in Fig. 2a and 3a are divided into two regions $r<1$, $s>0$ and $r>1$, $s<0$ by a vertical line passing through the point (1,0). The trajectories in the $\{r,s\}$ planes lying in the region $r<1$, $s>0$, a feature similar to the quintessence model(Q-model) of DE \cite{wuy}. The trajectories in the $\{r,s\}$ planes lying in the region $r>1$, $s<0$, a feature similar to the generalized Chaplygin gas (CG) model of DE \cite{sahni1,alam}. Here, we obtain a parabolic trajectory for both cases on $\lambda$. From Fig. 2a, for $\lambda=0.06$ we notice that the model behaves like $Q$-model for $0<(\zeta_0+\zeta_1)\le 0.46$, whereas for $(\zeta_0+\zeta_1)>0.46$ the model mimic like $CG$ model. The trajectories in both the regions converge to $\Lambda CDM$ model in late time of evolution. In the case $\lambda=-0.06$, we observe from Fig. 3a that all the $r-s$ trajectories lie in the region $(r<1, s>0)$ for $0<(\zeta_0+\zeta_1)\le 0.71$ which imply that the viscous new HDE model corresponds to $Q$ model while the trajectories lie in ($r>1, s<0$) region for $(\zeta_0+\zeta_1)>0.71$, {\it i.e.}, the model behaves like $CG$ model. In late time the viscous new HDE model approaches to $\Lambda CDM$. For some combinations like $(\zeta_0, \zeta_1)=(0.08,0.02)$, the trajectory starts in the vicinity of the $SCDM$ model and approaches to $\Lambda CDM$. Thus, we can conclude that for any value of $\lambda$ (either positive or negative), our viscous new HDE model in the framework of $f(R,T)$ theory mimic like $Q$ and $CG$ models for specific range of viscosity coefficients and in late time of evolution it always converges to $\Lambda CDM$ model.\\
\indent The $r-q$ trajectories in $r-q$ plane for positive and negative values of $\lambda$ and  for different combinations of ($\zeta_0 ,\zeta_1$) are shown in Figs 2b and 3b, respectively. Here, in both the figures stars represent the fixed point values $\{r,q\}=\{1,0.5\}$ for $SCDM$ model and $\{r,q\}=\{1,-1\}$ for Steady State ($SS$) model. The arrows represent the direction of the trajectories and the horizontal line at $r=1$ represents the time evolution of the $\Lambda CDM$ model. It can be observed that $q$ changes its sign from positive to negative with respect to time for $0<(\zeta_0+\zeta_1)\le 0.46$ in case of $\lambda=0.06$ and for $0<(\zeta_0+\zeta_1)\le 0.71$ in case $\lambda=-0.06$, which show the phase transition from decelerated phase to accelerated phase. For $(\zeta_0+\zeta_1)>0.46$ when $\lambda=0.06$ and $(\zeta_0+\zeta_1)>0.71$ when $\lambda=-0.06$, $q$ is always negative showing behavior of phantom. In the beginning this model behaves different from $\Lambda CDM$ model but in late time it behaves the same as $\Lambda CDM$ which converges to $SS$ model in late time evolution. In Fig.3b, we observe that for some small values of $(\zeta_0, \zeta_1)$, like $(\zeta_0, \zeta_1)=(0.08, 0.02)$, the $r-q$ trajectory starts in the neighbourhood of the $SCDM$ model. The present position of $\{r,s\}=\{r_0,s_0\}$ is indicated by dot in the plot. This means that the present viscous new HDE model is distinguishably different from the $\Lambda CDM$ model but in late time it converges to $SS$ model.\\
\indent The present viscous new HDE model can also be discriminated from the holographic dark energy model with event horizon as the infrared cutoff, in which the $r-s$ evolution starts from a region $r\sim 1$, $s\sim2/3$ and ends on the $\Lambda CDM$ point \cite{liu}. It can also be discriminated from Ricci dark energy model in which $(r,s)$ trajectory is a vertical segment, {\it i.e.}, $s$ is a constant during the evolution of the universe \cite{feng2}.
{\begin{center}
\begin{tabular}{cc}
\begin{minipage}{185pt}
\includegraphics[width=185pt]{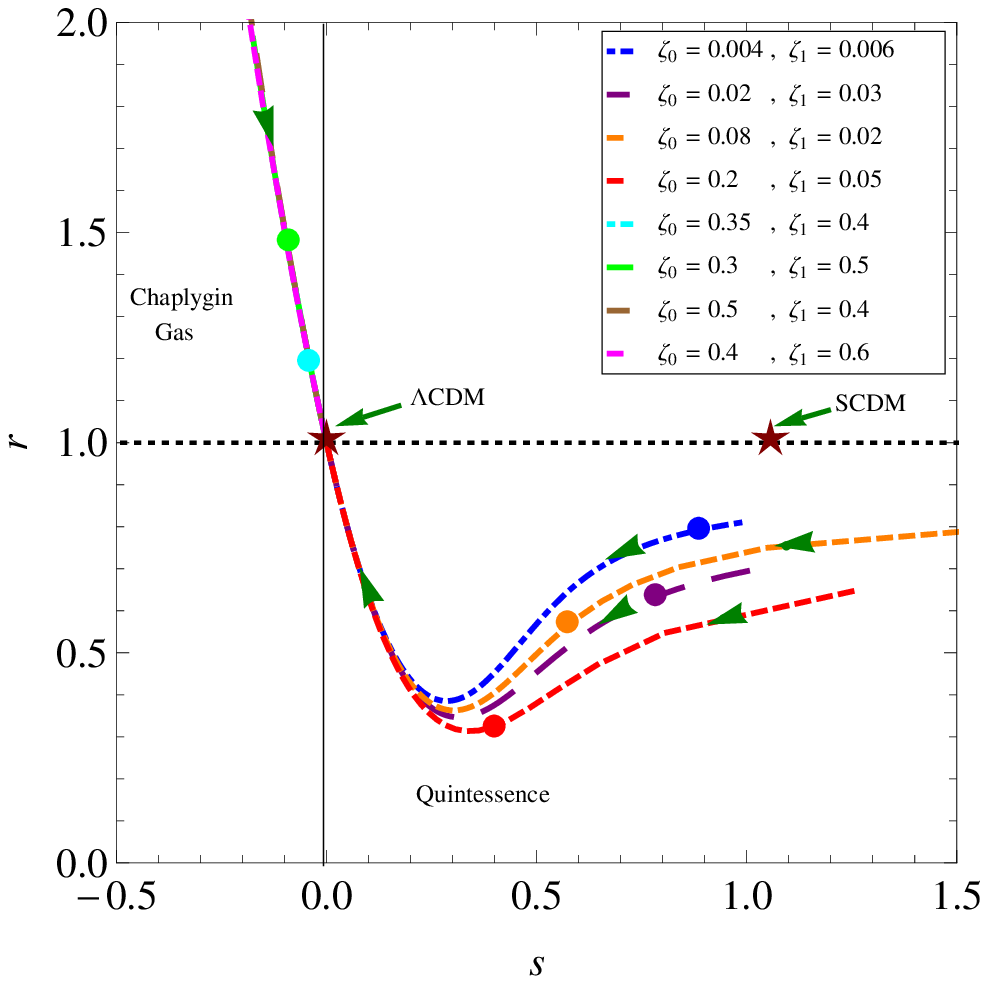}
{\footnotesize \textbf{Fig.3(a)} The $r-s$ trajectories are plotted in $r-s$ plane for different combinations of ($\zeta_0,\zeta_1$) with $\omega_d=-0.5$, $\alpha=0.8502$, $\beta=0.4817$ and $\lambda=-0.06$. The arrows represent the directions of the evolutions of statefinder diagnostic pair with time.}
\end{minipage}&\begin{minipage}{185pt}
\includegraphics[width=210pt]{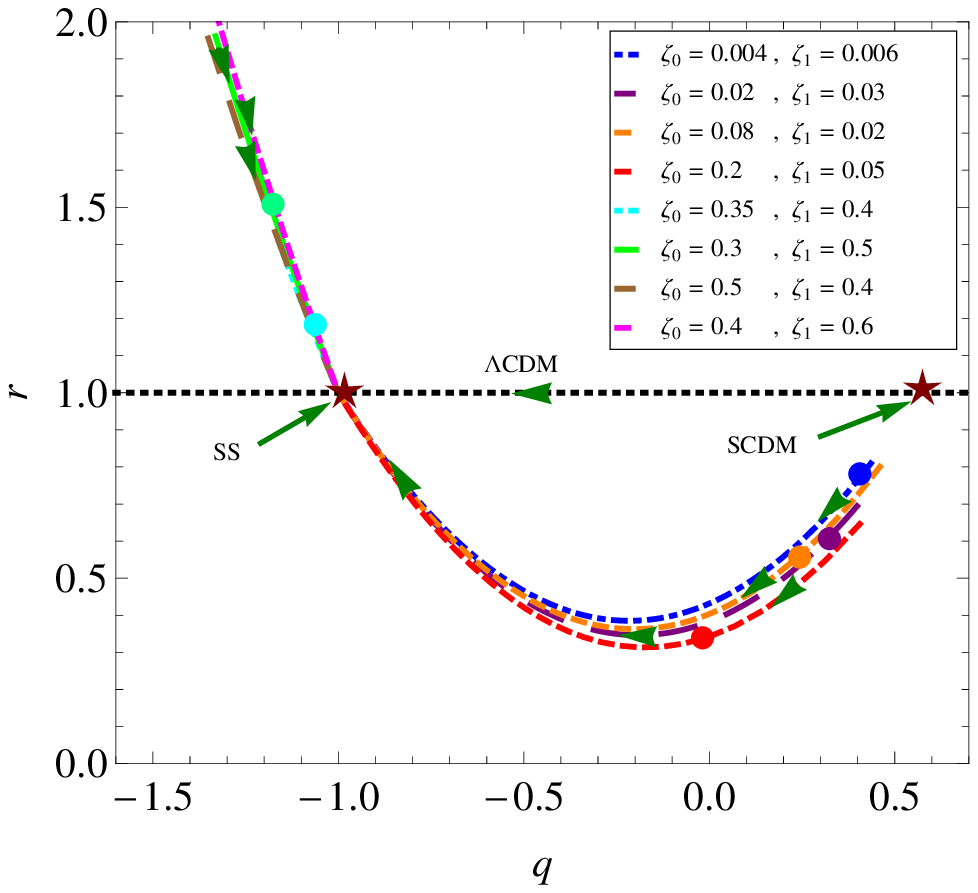}
{\footnotesize \textbf{Fig.3(b)} The $r-q$ trajectories are plotted in $r-q$ plane for different combinations of ($\zeta_0,\zeta_1$) with $\omega_d=-0.5$, $\alpha=0.8502$, $\beta=0.4817$ and $\lambda=-0.06$. The arrows represent the directions of the evolutions of statefinder diagnostic pair with time.}
\end{minipage}
\end{tabular}
\end{center}

\subsection{$\textit{Om}$ diagnostic}
\label{7}
Sahni {\it et al.} \cite{varun} introduced a new geometrical diagnostic known as $\textit{Om}$, which is a combination of the Hubble parameter and the cosmological redshift. Like the statefinder, $\textit{Om}$ depends only upon the expansion history of the universe. But, it's dependency only on the first derivative of the scale factor imply that it is easier to construct the $\textit{Om}$ as compare to statefinder parameters. It can discriminate dynamical dark energy models from $\Lambda CDM$, in a robust way, even if the value of the matter density is not precisely known. Secondly, it can provide a null test of $\Lambda CDM$ hypothesis, {\it i.e.}, $\textit{Om(z)}$ - $\Omega_{0m}=0$, if dark energy is a cosmological constant. $\textit{Om}$ has zero, negative and positive curvatures for $\Lambda CDM$, quintessence and phantom models, respectively. Many authors \cite{tong,luxu,haungt} have studied the DE models based on $\textit{Om(z)}$ diagnostic. Following Sahni {\it et al.}\cite{varun}, $\textit{Om(z)}$ for spatially flat universe is defined as
\begin{equation}
\textit{Om(z)} = \frac{\frac{H^2(z)}{H_0^2}-1}{(1+z)^3-1},
\end{equation}
where $H_0$ is the present value of the Hubble parameter. On substituting the required value of $H(z)$ from (29) into (42), we get the value of $\textit{Om(z)}$ as
\begin{equation}
\textit{Om(z)} = \frac{\frac{(1+2\lambda)^2}{[1+(1+2\lambda)(\alpha\omega_d-\zeta_1)]^2}\left[ \frac{\zeta_0}{H_0}+\Big\{\frac{ [1+(1+2\lambda)(\alpha \omega_d-\zeta_1)]}{(1+2\lambda)} - \frac{\zeta_0}{H_0}\Big\}(1+z)^{A[1+(1+2\lambda)(\alpha \omega_d-\zeta_1)]} \right]^2 - 1}{[(1+z)^3-1]}.
\end{equation}
{\begin{center}
\begin{tabular}{cc}
\begin{minipage}{190pt}
\includegraphics[width=190pt]{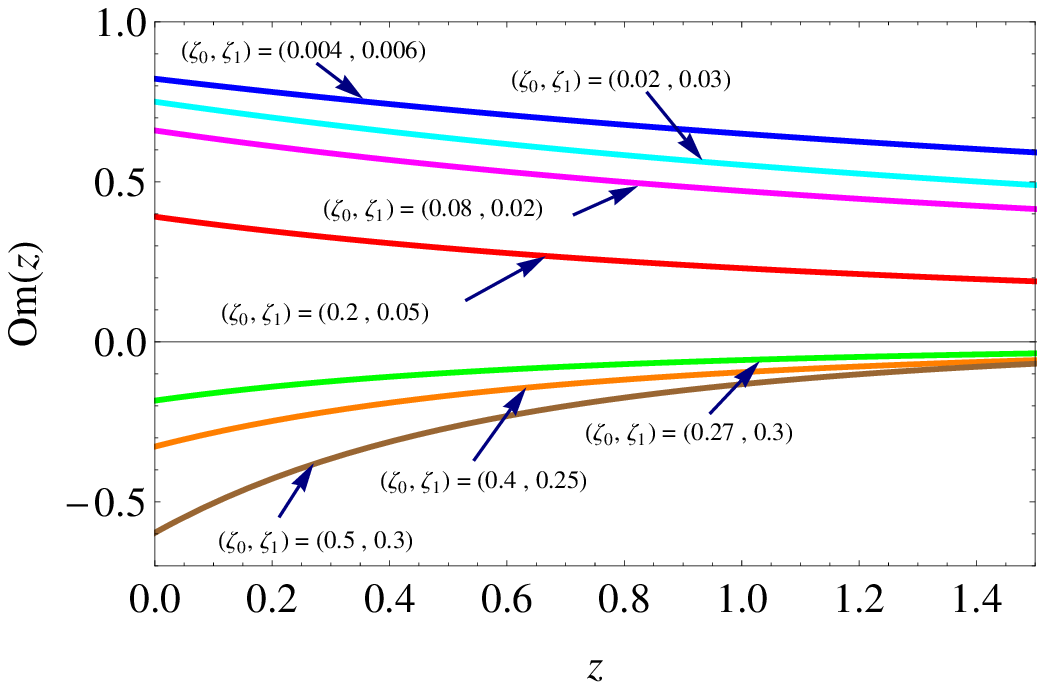}
{\footnotesize \textbf{Fig.4(a)} The $\textit{Om}-z$ trajectories are plotted in $\textit{Om}-z$ plane for different values of $\zeta_0$ and $\zeta_1$ with $\omega_d=-0.5$ and $\lambda=0.06$ along with the observational value of $\alpha=0.8502$ and $\beta=0.4817$.}
\end{minipage}&\begin{minipage}{190pt}
\includegraphics[width=190pt]{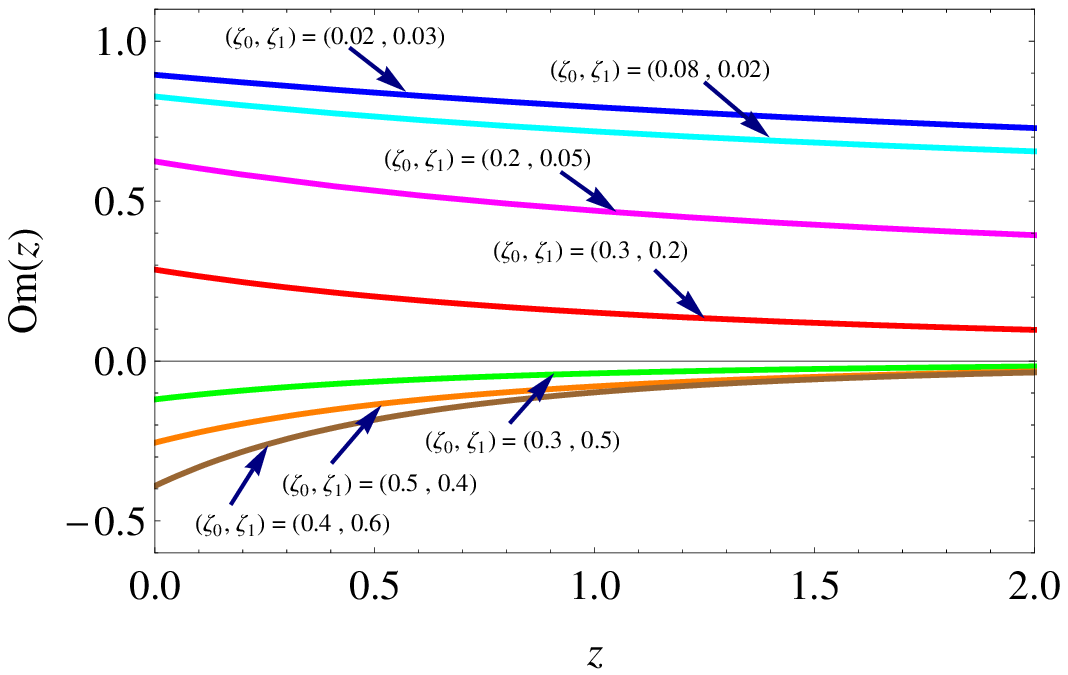}
{\footnotesize \textbf{Fig.4(b)} The $\textit{Om}-z$ trajectories are plotted in $\textit{Om}-z$ plane for different values of $\zeta_0$ and $\zeta_1$ with $\omega_d=-0.5$ and $\lambda=-0.06$ along with the observational value of $\alpha=0.8502$ and $\beta=0.4817$.}
\end{minipage}
\end{tabular}
\end{center}
In Fig. 4a, we plot the evolution of $\textit{Om(z)}$ against redshift $z$ corresponding to different values of a combination of ($\zeta_0, \zeta_1$), $\alpha=0.8502$, $\beta=0.4817$, $\omega_d=-0.5$ $H_0=1$ and $\lambda=0.06$. Similarly, in Fig. 4b, we plot the evolution of $\textit{Om(z)}$ against redshift $z$ corresponding to different values of a combination of ($\zeta_0, \zeta_1$), $\alpha=0.8502$, $\beta=0.4817$, $\omega_d=-0.5$ $H_0=1$ and $\lambda=-0.06$. The trajectory in Fig.4a is divided horizontally into two regions. In lower region, it may be seen that $\textit{Om(z)}$ decreases as $z$ decreases for $(\zeta_0+\zeta_1)> 0.46$, so positive slope of $\textit{Om(z)}$ suggests phantom $(\omega_d<-1)$ like behavior in the presence of viscosity with positive values of $\lambda=0.06$. However, in upper region, $\textit{Om(z)}$ increases as $z$ decreases for $0<(\zeta_0+\zeta_1)\le 0.46$, so negative slope of $\textit{Om(z)}$  indicating quintessence like behavior in the presence of viscosity.\\
\indent Similarly, the trajectory in Fig.4b is divided horizontally into two regions. In lower region, it may be seen that $\textit{Om(z)}$ decreases as $z$ decreases for $(\zeta_0+\zeta_1)> 0.71$, so positive slope of $\textit{Om(z)}$ suggests phantom $(\omega_d<-1)$ like behavior in the presence of viscosity with positive values of $\lambda=-0.06$. However, in upper region, $\textit{Om(z)}$ increases as $z$ decreases for $0<(\zeta_0+\zeta_1)\le 0.71$, so negative slope of $\textit{Om(z)}$  indicating quintessence like behavior in the presence of viscosity. In the late time of evolution when $z=-1$, we get $\textit{Om(z)}=1-\frac{(1+2\lambda)^2 \zeta_0^2}{[1+(1+2\lambda)(\alpha\omega_d-\zeta_1)]^2 H_0^2}$, which is the constant value of $\textit{Om(z)}$, {\it i.e.}, zero curvature. Thus, in late time the viscous new HDE corresponds to $\Lambda CDM$.

\subsection{Effective Equation of state parameter}

Let us discuss the bulk viscous effect on effective equation of state parameter, $\omega_{eff}$ which is given by
\begin{equation}
\omega_{eff}=\frac{p_d -3 \zeta H}{\rho_m+\rho_d},
\end{equation}
where $p_m=0$. On substituting the values of $p_d$, $\zeta$, $H$, $\rho_m$ and $\rho_d$ in Eq. (44), we get
\begin{equation}
\omega_{eff}=\frac{(2+3\lambda)\Big[\frac{\{(1+2\lambda)A\beta\omega_d-1\}
[(1+2\lambda)\zeta_0-\{1+(1+2\lambda)
(\alpha\omega_d-\zeta_1)\}H_0]}{e^{(1+2\lambda)A\zeta_0(t-t_0)}}-H_0 \Big]}{\Big[ \frac{\{(1+2\lambda)A\beta\omega_d-1\}
[(1+2\lambda)\zeta_0-\{1+(1+2\lambda)
(\alpha\omega_d-\zeta_1)\}H_0]\lambda}{e^{(1+2\lambda)A\zeta_0(t-t_0)}}+
(2+3\lambda+\alpha\lambda+2\alpha\lambda^2)H_0 \Big]}.
\end{equation}
The evolutions of $\omega_{eff}$ {\it versus} $t$ are shown in Figs. 5a and 5b for different pairs of $(\zeta_0, \zeta_1)$ in respect of $\lambda=0.06$ and $\lambda=-0.06$, respectively.  Figure 5a shows that the trajectories of $\omega_{eff}$ start from $\omega_{eff}>-1$ (it may also start from matter-dominated era) for small values and $\omega_{eff}<-1$ for large values of $(\zeta_0,\zeta_1)$ in respect of $\lambda=0.06$. As $t\rightarrow \infty$, $\omega_{eff}$ approaches to a constant for all values of $(\zeta_0, \zeta_1)$, {\it i.e.}, $\omega_{eff}\rightarrow -0.9745$. There is no $\omega_{eff}=-1$ crossing for small values of $(\zeta_0,\zeta_1)$ but for large values of $(\zeta_0,\zeta_1)$ it will cross the $\omega_{eff}=-1$. In Fig. 5b where we have $\lambda=-0.06$, we can observe the similar evolution of $\omega_{eff}$. However, $\omega_{eff}\rightarrow -1.0252$ in late times, {\it i.e.}, it crosses $\omega_{eff}=-1$ for small values of $(\zeta_0, \zeta_1)$ but it will not cross for large values of $(\zeta_0,\zeta_1)$. Thus, the effective equation of state parameter for both models shows consistency with the observational data given in Ref.\cite{ade}. We can say that the dark energy phenomena may be obtained in the presence of viscous fluid.
{\begin{center}
\begin{tabular}{cc}
\begin{minipage}{190pt}
\includegraphics[width=190pt]{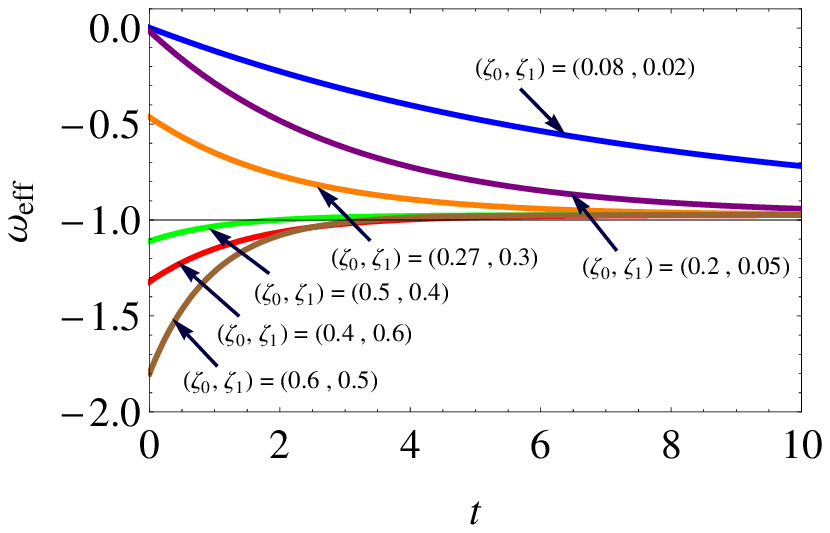}
{\footnotesize \textbf{Fig.5(a)} The evolution of $\omega_{eff}$ for different values of $(\zeta_0, \zeta_1)$ in respect of $\lambda=0.06$. We take $H_0=1$, $\omega_d=-0.5$, $\alpha=0.8502$ and $\beta=0.4817$.}
\end{minipage}&\begin{minipage}{190pt}
\includegraphics[width=190pt]{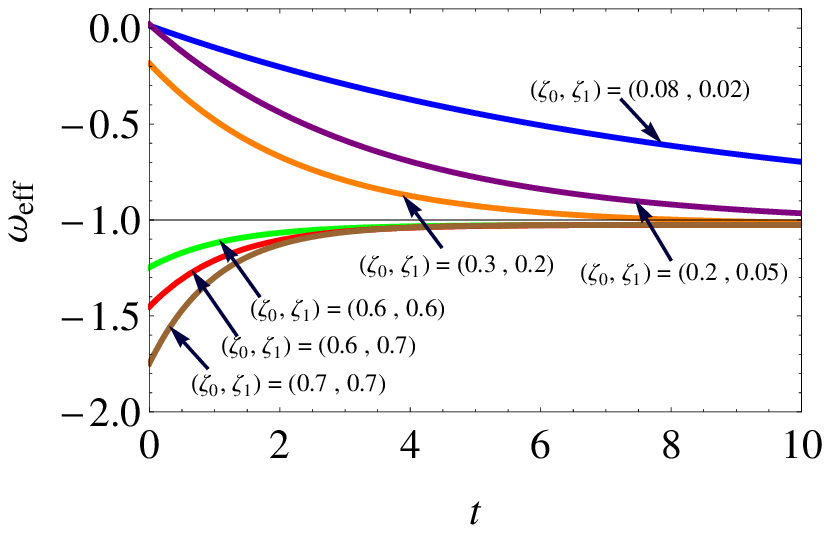}
{\footnotesize \textbf{Fig.5(b)} The evolution of $\omega_{eff}$ for different values of $(\zeta_0, \zeta_1)$ in respect of $\lambda=-0.06$. We take $H_0=1$, $\omega_d=-0.5$, $\alpha=0.8502$ and $\beta=0.4817$.}
\end{minipage}
\end{tabular}
\end{center}

\subsection{Entropy and second law of thermodynamics}
\label{8}
The local entropy production for a fluid on a flat FRW spacetime is expressed as \cite{weinberg}
\begin{equation}
T\triangledown_{\nu}s^{\nu}=\zeta (\triangledown_{\nu}u^{\nu})^2=9H^2\zeta
\end{equation}
\noindent where $T$ is the temperature, $\triangledown_{\nu}s^{\nu}$ is the rate at which entropy is being generated in unit volume, and $\zeta$ is the total bulk viscosity.\\
\indent The second law of thermodynamics can be stated as
\begin{equation}
T\triangledown_{\nu}s^{\nu}\geq 0
\end{equation}
Since the Hubble parameter $H$ is positive in an expanding universe, then $\zeta$ has to be positive in order to preserve the validity of the second law of thermodynamics. Thus, equation (46) implies that
\begin{equation}
\zeta\geq 0.
\end{equation}
Thus, for the present model the inequality (48) can be written as
\begin{equation}
\zeta=\zeta_0+\zeta_1H\geq 0.
\end{equation}
Using (29), we find the expression for the total bulk viscosity $\zeta(a)$ as
\begin{eqnarray}
\zeta(a)&=&\zeta_0+\zeta_1 \Bigg[ \frac{(1+2\lambda)H_0}{[1+(1+2\lambda)(\alpha\omega_d-\zeta_1)]}\Bigg( \frac{\zeta_0}{H_0}+\left\{ \frac{[1+(1+2\lambda)(\alpha\omega_d-\zeta_1)]}{(1+2\lambda)}-\frac{\zeta_0}{H_0} \right\}\nonumber\\&&+\;\;a^{-A\{1+(1+2\lambda)(\alpha\omega_d-\zeta_1)\}}\Bigg)\Bigg].
\end{eqnarray}
\indent The value of the scale factor, at which the transition of the total bulk viscosity between negative to positive values happen, is
\begin{equation}
a_{np}=\left[ \frac{\{1+(1+2\lambda)\alpha\omega_d\}\zeta_0}{\{(1+2\lambda)(\zeta_0+\zeta_1 H_0-\alpha\omega_d)-1\}\zeta_1} \right]^{-\frac{1}{A[1+(1+2\lambda)(\alpha\omega_d-\zeta_1)]}},
\end{equation}
where, the subscript ``np" stands for ``negative to positive" values. In late time of evolution, {\it i.e.}, at $a\rightarrow \infty$ the total bulk viscosity is $\zeta(a) = \frac{\{1+(1+2\lambda)\alpha\omega_d\} \zeta_0}{\{1+(1+2\lambda)(\alpha\omega_d-\zeta_1)\}}$.

{\begin{center}
\begin{tabular}{cc}
\begin{minipage}{190pt}
\includegraphics[width=190pt]{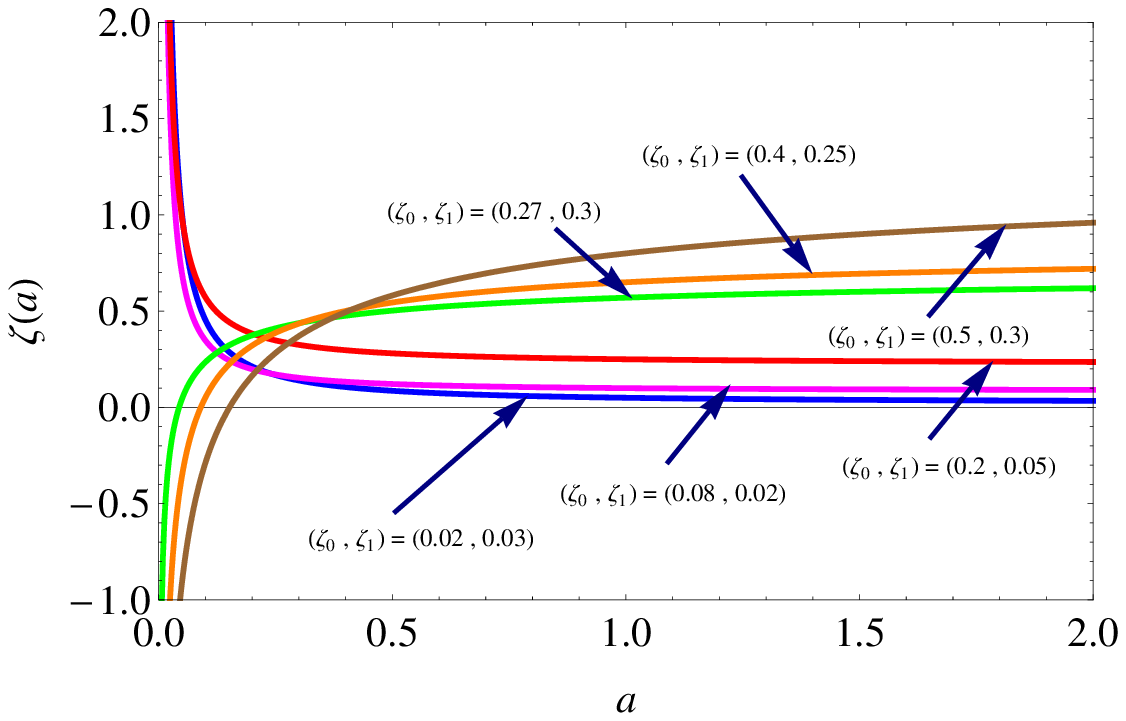}
{\footnotesize \textbf{Fig.6(a)} The evolution of $\zeta(a)$  for different combination of $\zeta_0$ and $\zeta_1$ with $\omega_d=-0.5$, $\lambda=0.06$, $\alpha=0.8502$ and $\beta=0.4817$.}
\end{minipage}&\begin{minipage}{190pt}
\includegraphics[width=190pt]{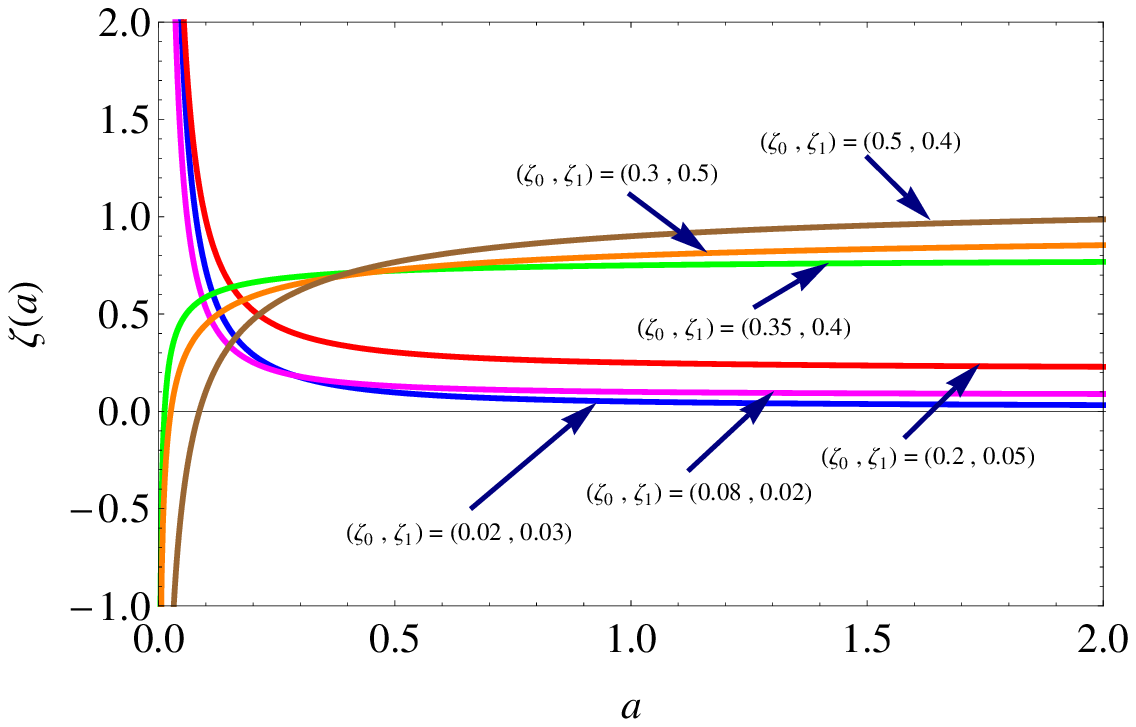}
{\footnotesize \textbf{Fig.6(b)} The evolution of $\zeta(a)$  for different combination of $\zeta_0$ and $\zeta_1$ with $\omega_d=-0.5$, $\lambda=-0.06$, $\alpha=0.8502$ and $\beta=0.4817$.}
\end{minipage}
\end{tabular}
\end{center}
Figures 6a and 6b represent the evolution of the total bulk viscous coefficient $\zeta$ with respect to the scale factor for the different combinations of the $(\zeta_0, \zeta_1)$ with positive ($\lambda =0.06$) and negative ($\lambda=-0.06$) values of the coupling parameter, respectively. The figures 6a and 6b show that the total bulk viscosity is positive throughout the evolution of the universe for $0<(\zeta_0+\zeta_1)\le 0.46$ with $\lambda=0.06$, and $0<(\zeta_0+\zeta_1)\le 0.71$ with $\lambda=-0.06$, respectively. These curves have been shown above the line (0,0). Therefore, the model does not violate the entropy law for these ranges. The figures also show that the total bulk viscous coefficient is evolving from negative to positive value for $(\zeta_0+\zeta_1)\ge 0.46$ and $(\zeta_0+\zeta_1)\ge 0.71$, respectively. Thus, the rate of entropy production is negative for large values of $(\zeta_0+\zeta_1)$ in early epoch and positive in the later epoch. Hence the entropy law violates in the early epoch and obeys in the later epoch for these values of $(\zeta_0+\zeta_1)$. \\
\indent In an absolute way the status of the second law of thermodynamics should be considered along with the accounting of the entropy generation from the horizon. In those circumstances, the second law becomes the generalised second law (GSL) of thermodynamics, according to which the total sum of the entropies of the fluid components of the universe plus that of the horizon entropy should never decrease \cite{gib,mathew}. In the present model this means the rate of change of complete entropy along with horizon must be greater than or equal to $0$. Mathew {\it et al.}\cite{mathew} have discussed the status of GSL for flat FRW universe with matter and cosmological vacuum. Karami {\it et al.} \cite{karami} have discussed the status of GSL in a flat universe with viscous dark energy and have shown that the GSL is valid in FRW universe with apparent horizon as the boundary. Many authors \cite{sasi,mr} have verified the status of GSL with apparent horizon as the boundary. The status of GSL in the modified gravity theories have also been discussed by many authors \cite{shao,zubair1,zubair2,mome} with the apparent horizon as the boundary.\\
\indent Let us verify the GSL of thermodynamics for this model. As stated above we consider the apparent horizon as the boundary of the universe. Then, the GSL can be stated as
\begin{equation}
\frac{d}{dt} (S_{tot})=\frac{d}{dt} (S_m+ S_d+ S_p+{\tilde{S_h}})\ge 0,
\end{equation}
where $S_m$, $S_d$, $S_p$ and $\tilde{S_h}$ are the entropies of the dark matter, dark energy, the entropy production and the entropy of the apparent horizon, respectively.\\
\indent From the first law of thermodynamics the change of entropy of the viscous matter inside the apparent horizon can be obtained using Gibbs equation
\begin{equation}
T_i dS_i = d(\rho_i V) +p_i dV,
\end{equation}
\noindent where $T_i$ is the temperature and $S_i=S_m+S_d$, is the sum of the entropies of the dark matter and dark energy, $\rho_i$ represents the sum of the densities of the dark matter and dark energy, $V= \frac{4\pi \tilde{r_h}}{3}$ is the volume of the apparent horizon with $\tilde{r_h}$ as a radius of the apparent horizon and $p_i=\tilde{p_d}=p_d-3\zeta H$ as the effective pressure.\\
\indent In the present viscous model the above Gibbs equation modifies to \cite{zubair1,zubair2,mome}
\begin{equation}
T_i dS_i = d(\rho_i V) +p_i dV - T_i dS_p.
\end{equation}
\indent The radius $\tilde{r_h}$ of the apparent horizon for a flat FRW universe, is defined as \cite{mome}
\begin{equation}
\tilde{r_h}=\frac{1}{H}.
\end{equation}
In the $f(R,T)$ theories the entropy associated with the apparent horizon is defined as
\begin{equation}
\tilde{S_h}=\frac{\tilde{A_h}}{4}\frac{f_R(R,T)}{\left(\frac{1}{8\pi}+\frac{f_T(R,T)}
{8\pi}\right)}=\frac{2\pi \tilde{A_h}}{(1+\lambda)}=\frac{8\pi^2\tilde{r_h}^2}{(1+\lambda)},
\end{equation}
where $\tilde{A_h}=4\pi\tilde{r_h}^2$ is the area of the apparent horizon. In this paper, we have taken $8\pi G=1$. Taking the derivative of Eq. (56) with respect to time $t$, we get
\begin{equation}
\dot{\tilde{S_h}}=\frac{16\pi^2\tilde{r_h}\dot{\tilde{r_h}}}{(1+\lambda)}=
\frac{16\pi^2}{(1+\lambda)H}\left(-\frac{\dot H}{H^2}\right).
\end{equation}
\indent Now, from (54) we have
\begin{equation}
T_i(\dot S_m+\dot S_d+\dot S_p)=4\pi \tilde{r_h}^2(\dot{\tilde{r_h}}-1)\{\rho_m+(1+\omega_d)\rho_d-3\zeta_0H-3\zeta_1 H^2\}.
\end{equation}
Under the thermal equilibrium conditions between the fluids and the horizon, we have $T_i=\tilde{T_h}$. We take the temperature $\tilde{T_h}=\frac{H}{2\pi}$, which is equal to the Hawking temperature of the horizon with the assumption that the fluid within the horizon is in equilibrium with the horizon, so there is no effective flow of the fluid toward the horizon\cite{mathew1}. Now, Eq. (58) become as
\begin{equation}
\dot S_m+\dot S_d+\dot S_p=\frac{8\pi^2}{H^3}\left(-\frac{\dot H}{H^2}-1\right)\{\rho_m+(1+\omega_d)\rho_d-3\zeta_0H-3\zeta_1 H^2\}.
\end{equation}
Now, using the definition of deceleration parameter $q=\left(-\frac{\dot H}{H^2}-1\right)$ and  Eqs. (57) and (59) into (52), we get the change of the sum of all the entropies as
\begin{eqnarray}
\dot S_{tot}&&=\dot S_m+\dot S_d+\dot S_p+\dot{\tilde{S_h}}\nonumber\\&&=\frac{8\pi^2 q}{H^3}\{\rho_m+(1+\omega_d)\rho_d-3\zeta_0H-3\zeta_1 H^2\}
+\frac{16\pi^2(q+1)}{(1+\lambda)H}.
\end{eqnarray}
On substituting the required values in above equation and simplifying, we get the change in total entropy as
\begin{eqnarray}
\dot S_{tot}&= \left[\frac{16\pi^2[\{1+(1+2\lambda)(\alpha\omega_d-\zeta_1)\}H_0-(1+2\lambda)
\zeta_0]\left\{1+\frac{\{1+(1+2\lambda)(\alpha\omega_d-\zeta_1)\}H_0\left( e^{(1+2\lambda) A\zeta_0(t-t_0)}-1\right)}{(1+2\lambda)\zeta_0} \right\}}{H_0^2 e^{2(1+2\lambda)A\zeta_0(t-t_0)}}\right]\nonumber\\&\times \left[\frac{A}
{(1+\lambda)}+\frac{3\{1-(1+2\lambda)A\beta\omega_d\}}{(2+3\lambda)H_0}
\left\{\frac{[\{1+(1+2\lambda)(\alpha\omega_d-\zeta_1)\}H_0-(1+2\lambda)\zeta_0]A}
{e^{(1+2\lambda) A\zeta_0(t-t_0)}}-H_0\right\} \right].
\end{eqnarray}
From Eq. (61), it is noticed that the rate of change of entropy can not clearly state either the change in entropy is greater than and equal to zero. We plot the evolution of $\dot{S}$ in Figs. 7a and 7b, respectively for positive and negative values of the $\lambda$ ({\it e.g.,} $\lambda=0.06$, $\lambda=-0.06$) taking $\alpha=0.8502$, $\beta=0.4817$, $\omega_d=-0.5$, $H_0=1$ and $t_0=1$ for different values of viscosity coefficients $(\zeta_0, \zeta_1)$. Figure 7a shows that the GSL is always valid for $\zeta_0>0$ and $0<\zeta_1\le 0.85$ whereas it holds for $\zeta_0>0$ and $0<\zeta_1\le 1$ in Fig. 7b. We also notice that in both the cases, the total entropy corresponds to zero in late time of the evolution.

{\begin{center}
\begin{tabular}{cc}
\begin{minipage}{190pt}
\includegraphics[width=190pt]{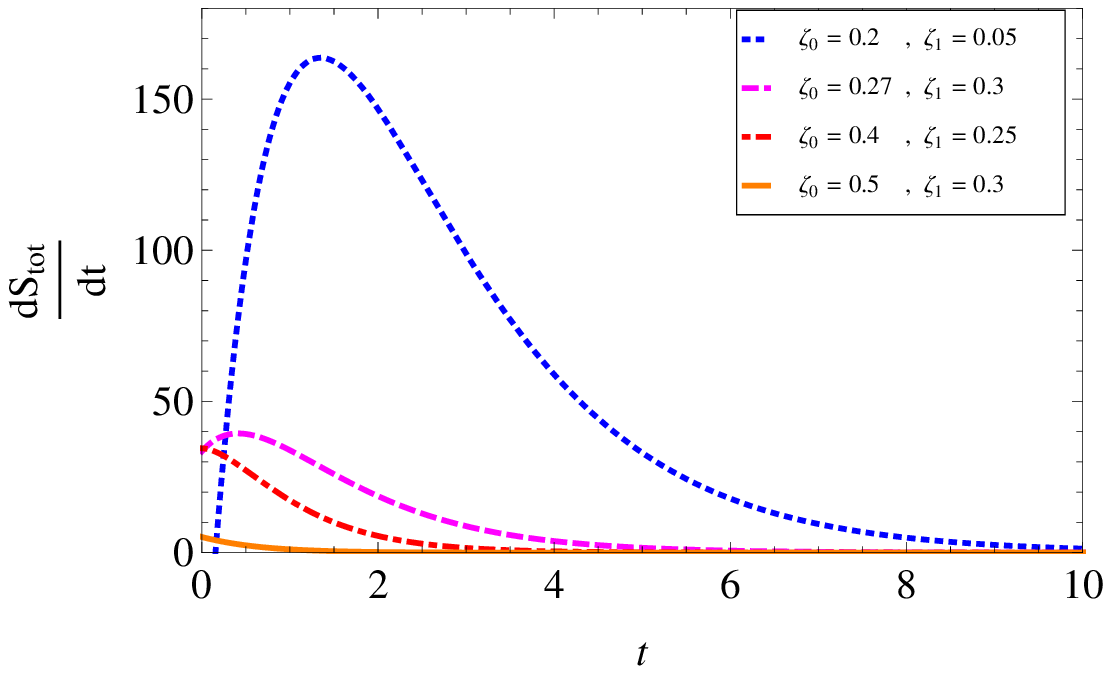}
{\footnotesize \textbf{Fig.7(a)} The change in total entropy with respect to $t$ for different combination of $\zeta_0$ and $\zeta_1$ with $\omega_d=-0.5$, $\lambda=0.06$, $\alpha=0.8502$ and $\beta=0.4817$.}
\end{minipage}&\begin{minipage}{190pt}
\includegraphics[width=190pt]{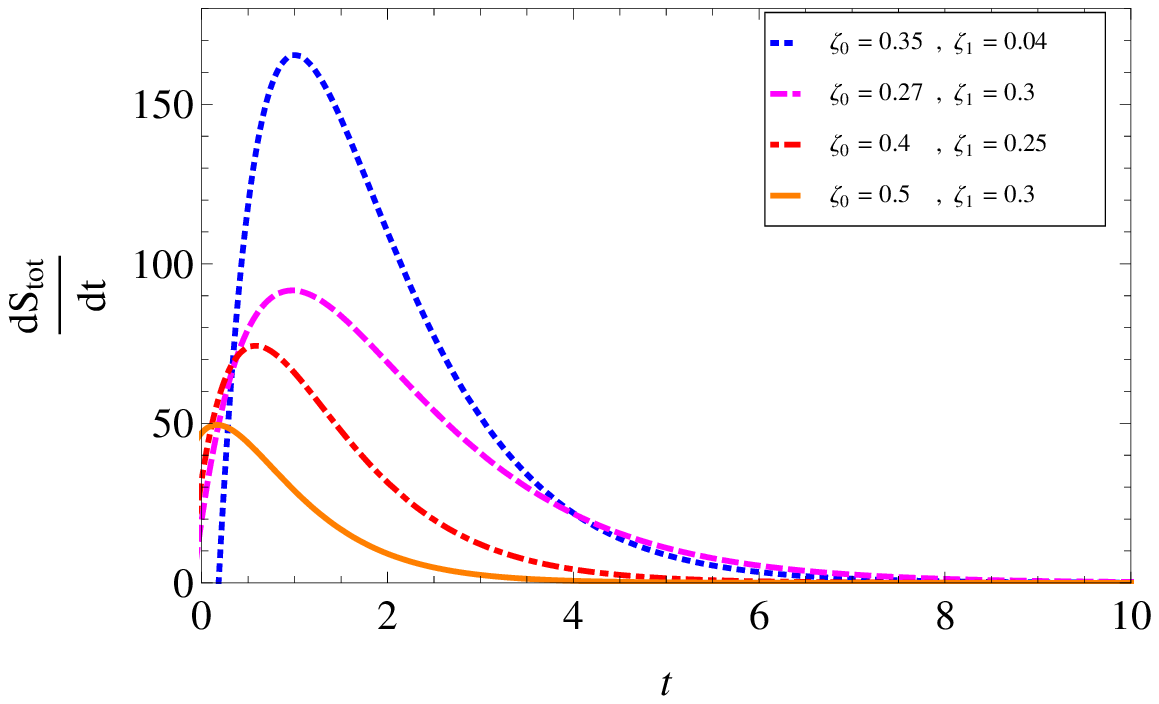}
{\footnotesize \textbf{Fig.7(b)} The change in total entropy with respect to $t$ for different combination of $\zeta_0$ and $\zeta_1$ with $\omega_d=-0.5$, $\lambda=-0.06$, $\alpha=0.8502$ and $\beta=0.4817$.}
\end{minipage}
\end{tabular}
\end{center}

\section{Conclusion}

In this paper, we have explored the bulk viscous new HDE model in modified $f(R,T)$ gravity with bulk viscosity of the form $\zeta=\zeta_0+\zeta_1 H$. We have considered that the universe is filled with the pressureless dark matter and the new HDE proposed by Granda and Oliveros \cite{go}. We have analyzed the different possible phases of the universe according to the dimensionless bulk viscous parameters $\zeta_0$ and $\zeta_1$. We have calculated the scale factor, deceleration parameter, statefinder and $\textit{Om}$ diagnostics, effective equation of state parameter, entropy and generalized second law of thermodynamics to observe the evolution of the universe. In what follows we summarize the result.\\
\indent The field equations of viscous new HDE model in modified $f(R,T)$ gravity have been solved for a most general bulk viscous coefficient. The scale factor has been obtained in the form of exponential which describes the phase transition of the universe. We have analyzed the behavior of the scale factor for all possible combination of ($\zeta_0,\zeta_1$) and model parameter $\lambda$. We have obtained the cosmic time when the Big Bang happens. In early time the scale factor is approximated by power-law expansion which shows that the model expands with decelerated rate in early time. In late time the scale factor behaves as a de Sitter like expansion, {\it i.e}, expands with accelerated rate. This shows that the scale factor at the respective limits has an earlier decelerated phase followed by an accelerated phase in the later stage of the evolution. We have calculated the transition time between decelerated to accelerated epoch. We have expressed Hubble parameter in terms of scale factor to get the transition between decelerated to accelerated epoch in terms of scale factor $a_T$ and redshift $z_T$. Thus, for the present time transition from deceleration to acceleration takes place at $\zeta_0+\zeta_1= 0.096$ for positive values of $\lambda = 0.06$ and for negative value of $\lambda = -0.06$, we get $\zeta_0+\zeta_1= 0.218$. A plot of the evolution of the scale factor is given in Fig. 1. for difference values of a combination of $\zeta_0$ and $\zeta_1$ and positive value of $\lambda$. For $0 < (\zeta_0+\zeta_1)<0.096$, the scale factor has an earlier deceleration phase followed by an acceleration phase in late time. For $(\zeta_0+\zeta_1) > 0.096$, the transition takes place in past of the universe. Similar behavior can be observed for the negative value of $\lambda$.\\
\indent We have also studied the evolution of the deceleration parameter $q$. We have obtained time-dependent deceleration parameter. As $(t-t_0) \rightarrow \infty$, we get $q \rightarrow -1$, which shows that the model accelerates in late time. We have observed that for $\zeta_0+\zeta_1= 0.096$, we have $q_0=0$. Therefore, the transition into acceleration phase would occur at present time. We have discussed the behavior of $q$ and corresponding evolution for different ranges of coupling parameter $\lambda$ and constraints on $(\zeta_0 + H_0\zeta_1)$. The results are summarized in tables 1-3 for $\omega_d=-0.5$, $\omega_d=-1$ and $\omega_d=-1.1414$, and model parameters $\alpha=0.8502$, $\beta=0.4817$. The tables show that the deceleration or acceleration or their transition depend on the values of $\zeta_0$, $\zeta_1$ and $\lambda$. The behavior of $q$ has already been discussed in detail in subsection 3.2 .\\
\indent We have discussed two geometrical diagnostics, namely statefinder and $\textit{Om}$ to observe the discrimination with the other existing DE models. Let us first discuss about the statefinder diagnostic for viscous new HDE model. In the late time as $(t-t_0) \rightarrow \infty$, the statefinder pair $\{r,s\}$ for the viscous new HDE model tends to $\{1,0\}$, a value corresponding to $\Lambda CDM$ model. We have plotted the trajectory of statefinder pair $\{r,s\}$ in $r-s$ plane for different combinations of $(\zeta_0, \zeta_1)$ as shown in Fig.2a and Fig. 3a for positive and negative values of coupling parameter $\lambda$, respectively. It has been observed that for any value of $\lambda$ (either positive or negative), our viscous new HDE model in the framework of $f(R,T)$ theory mimics like $Q$ and $CG$ models for specific range of viscosity coefficients and in late time of evolution it always converges to $\Lambda CDM$ model. \\
\indent We have also plotted the trajectory of $\{r,q\}$ for different combinations of $(\zeta_0, \zeta_1)$ and positive and negative values of $\lambda$ as shown in Fig. 2b and Fig. 3b. It can be observed that $q$ changes its sign from positive to negative with respect to time for $0<(\zeta_0+\zeta_1)\le 0.46$ in case of $\lambda=0.06$ and for $0<(\zeta_0+\zeta_1)\le 0.71$ in case $\lambda=-0.06$, which show the phase transition from decelerated phase to accelerated phase. For $(\zeta_0+\zeta_1)>0.46$ when $\lambda=0.06$ and $(\zeta_0+\zeta_1)>0.71$ when $\lambda=-0.06$, $q$ is always negative showing behavior of phantom. Thus, in the beginning this model behaves different from $\Lambda CDM$ model but in late time it behaves the same as $\Lambda CDM$ which converges to $SS$ model in late time evolution. The present viscous new HDE model can also be discriminated from the holographic dark energy model with event horizon as the infrared cutoff, in which the $r-s$ evolution starts from a region $r\sim 1$, $s\sim2/3$ and ends on the $\Lambda CDM$ point \cite{liu}. It can also be discriminated from Ricci dark energy model in which $r-s$ trajectory is a vertical segment, i.e., $s$ is a constant during the evolution of the universe \cite{feng2}.\\
\indent The second geometrical diagnostic, namely $\textit{Om}$ has been carried out in section 3.4 . We have plotted the $\textit{Om(z)}$ versus redshift $z$ trajectories for different combinations of $(\zeta_0, \zeta_1)$ in respect of positive and negative values of $\lambda$ as shown in Fig.4a and Fig. 4b, respectively. We have observed two types of trajectory, one has the positive curvature which suggests the phantom like behavior and second has the negative curvature which suggests the quintessence like behavior.\\
\indent We have calculated the effective EoS parameter for this model and analyzed the evolution of it graphically for different suitable values of viscosity coefficients. We have observed that  $\omega_{eff}\rightarrow$ -0.9745 and -1.0252, respectively as $t\rightarrow \infty$ in case positive and negative values of $\lambda$ which is very close to -0.93 predicted in the Ref.\cite{kom}.\\
\indent At the end, we have discussed the entropy and second law of thermodynamics for viscous new HDE model. Figures 6a and 6b plot the evolution of the total bulk viscous coefficient. It has been observed that the total bulk viscosity is positive throughout the evolution of the universe for $0<(\zeta_0+\zeta_1)\le 0.46$ with $\lambda=0.06$, and $0<(\zeta_0+\zeta_1)\le 0.71$ with $\lambda=-0.06$, respectively. Therefore, the model does not violate the entropy law for these ranges. The figures also show that $\zeta(a)$ is evolving from negative to positive value for $(\zeta_0+\zeta_1)> 0.46$ and $(\zeta_0+\zeta_1)> 0.71$, respectively. Thus, the rate of entropy production is negative for large values of $(\zeta_0+\zeta_1)$ in early epoch and positive in the later epoch. Hence the entropy law violates in the early epoch and obeys in the later epoch for these values of $(\zeta_0+\zeta_1)$. We have also studied the generalized second law of thermodynamics. It has been observed that it is always valid for $\zeta_0>0$ and $0<\zeta_1\le 0.85$ when $\lambda$ is positive whereas it holds for $\zeta_0>0$ and $0<\zeta_1\le 1$ when $\lambda$ is negative which are shown in Figs. 7a and 7b, respectively. We have also noticed that in both the cases, the total entropy corresponds to zero in late time of the evolution.\\
\indent Let us Summarize the results with the outcome of this work. In cosmology, the idea of viscous dark energy models has been presented in different ways to understand the evolution of the universe. The notion of bulk viscosity has also been extensively studied in modified theories. This paper has explored the behavior of viscosity by considering dust matter and new HDE in the background of modified $f(R,T)$ gravity. The above investigations show that this model predicts an early deceleration followed by late time acceleration. We can conclude that the dark energy era may be obtained in the presence of bulk viscous fluid. \\\\

\noindent \textbf{Acknowledgements}\\
One of the author MS would like to thank to University Grant Commission, India for providing Senior Research Fellowship (SRF) under UGC-NET scholarship.\\

\end{document}